\documentclass{pasa}%
\usepackage{graphicx}
\usepackage{amsmath}	% Advanced maths commands
\usepackage{amssymb}	% Extra maths symbols
\usepackage{siunitx}
\usepackage{pdflscape}

\newcommand{\src}[1]{\vspace*{4pt}\noindent {\bf #1}}
\newcommand{\software}{\texttt}
\newcommand{\askapsoft}{\software{ASKAPSoft}}
\newcommand{\trap}{\software{TraP}}
\newcommand{\casda}{{CASDA}}
\newcommand{\selavy}{\software{Selavy}}
\newcommand{\vasttools}{\software{VAST Tools}}
\newcommand{\wise}{\textit{WISE}}
\newcommand{\gaia}{\textit{Gaia}}
\newcommand{\red}[1]{{#1}} 

\title[ASKAP VAST Pilot Survey]{The ASKAP Variables and Slow Transients (VAST) Pilot Survey}

\author[Tara Murphy et al.]{Tara Murphy,$^{1,2}$\thanks{tara.murphy@sydney.edu.au}
David L. Kaplan,$^3$ 
Adam J. Stewart,$^1$
Andrew O'Brien,$^3$
Emil Lenc,$^4$
Sergio Pintaldi,$^5$
Joshua Pritchard,$^{1,2,4}$
Dougal Dobie,$^{2,6}$
Archibald Fox,$^{1}$
James K. Leung,$^{1,2,4}$
Tao An,$^{7}$
Martin E. Bell,$^{8}$
Jess W. Broderick,$^{9}$
Shami Chatterjee,$^{10}$
Shi Dai,$^{11}$
Daniele d'Antonio,$^{4,8}$
J. Gerry Doyle,$^{12}$
B. M. Gaensler,$^{13}$
George Heald,$^{14}$
Assaf Horesh,$^{15}$
Megan L. Jones,$^{3}$
David McConnell,$^{4}$
Vanessa A. Moss,$^{4,1}$
Wasim Raja,$^{4}$
Gavin Ramsay,$^{11}$
Stuart Ryder,$^{16,17}$
Elaine M. Sadler,$^{1,4}$
Gregory R. Sivakoff,$^{18}$ 
Yuanming Wang,$^{1,2,4}$ 
Ziteng Wang,$^{1,2,4}$
Michael S. Wheatland,$^{1}$
Matthew Whiting,$^{4}$
James R. Allison,$^{19,4}$
C. S. Anderson,$^{20,4}$
Lewis Ball,$^{4,21}$
K. Bannister,$^{4}$
D. C.-J. Bock,$^{4}$
R. Bolton,$^{4}$
J. D. Bunton,$^{4}$
R. Chekkala,$^{4}$
A. P. Chippendale,$^{4}$
F. R. Cooray,$^{4}$
N. Gupta,$^{22}$
D. B. Hayman,$^{4}$
K. Jeganathan,$^{4}$
B. Koribalski,$^{4,11}$
K. Lee-Waddell,$^{23,4}$
Elizabeth K. Mahony,$^{4}$
J. Marvil,$^{24}$
N.~M.\ McClure-Griffiths,$^{25}$
P. Mirtschin,$^{4}$
A. Ng,$^{4}$ 
S. Pearce,$^{4}$
C. Phillips$^{4}$ and
M. A. Voronkov$^{4}$
\affil{$^1$Sydney Institute for Astronomy, School of Physics, The University of Sydney, NSW 2006, Australia}
\affil{$^2$ARC Centre of Excellence for Gravitational Wave Discovery (OzGrav), Hawthorn, Victoria, Australia}
\affil{$^3$Department of Physics, University of Wisconsin-Milwaukee, P.O. Box 413, Milwaukee, WI 53201, USA}
\affil{$^4$CSIRO, Space and Astronomy, PO Box 76, Epping, NSW 1710, Australia}
\affil{$^5$Sydney Informatics Hub, The University of Sydney, NSW 2008, Australia}
\affil{$^{6}$Centre for Astrophysics and Supercomputing, Swinburne University of Technology, Hawthorn, Victoria, Australia} 
\affil{$^{7}$Shanghai Astronomical Observatory, the Chinese Academy of Sciences, 80 Nandan Road, Shanghai 200030, China}
\affil{$^8$University of Technology Sydney, 15 Broadway, Ultimo NSW 2007, Australia}
\affil{$^{9}$International Centre for Radio Astronomy Research, Curtin University, GPO Box U1987, Perth, WA 6845, Australia}
\affil{$^{10}$Cornell Center for Astrophysics and Planetary Science, Ithaca, NY 14853, USA}
\affil{$^{11}$School of Science, Western Sydney University, Locked Bag 1797, Penrith South DC, NSW 2751, Australia}
\affil{$^{12}$Armagh Observatory and Planetarium, College Hill, Armagh, BT61 9DG, N. Ireland}
\affil{$^{13}$Dunlap Institute for Astronomy and Astrophysics, University of Toronto, 50 St. George St., Toronto, ON M5S 3H4, Canada; David A. Dunlap Department of Astronomy and Astrophysics, University of Toronto, 50 St. George St., Toronto, ON M5S 3H4, Canada}
\affil{$^{14}$CSIRO, Space and Astronomy, PO Box 1130, Bentley WA 6102, Australia}
\affil{$^{15}$Racah Institute of Physics, The Hebrew University of Jerusalem, Jerusalem 91904, Israel}
\affil{$^{16}$Department of Physics and Astronomy, Macquarie University, Sydney NSW 2109, Australia}
\affil{$^{17}$Astronomy, Astrophysics and Astrophotonics Research Centre, Macquarie University, Sydney, NSW 2109, Australia}
\affil{$^{18}$Department of Physics, University of Alberta, CCIS 4-181, Edmonton, AB T6G 2E1, Canada}
\affil{$^{19}$Sub-Dept. of Astrophysics, Department of Physics, University of Oxford, Denys Wilkinson Building, Keble Rd., Oxford, OX1 3RH, UK}
\affil{$^{20}$Jansky Fellow of the National Radio Astronomy Observatory, P. O. Box 0,Socorro, NM 87801, USA}
\affil{$^{21}$SKA Observatory,  Jodrell Bank, Lower Withington, Macclesfield, Cheshire SK11 9FT, UK}
\affil{$^{22}$Inter-University Centre for Astronomy and Astrophysics, Post Bag 4, Ganeshkhind, Pune University Campus, Pune 411 007, India}
\affil{$^{23}$International Centre for Radio Astronomy Research (ICRAR), The University of Western Australia, 35 Stirling Hwy, Crawley, WA, 6009, Australia}
\affil{$^{24}$National Radio Astronomy Observatory, P.O. Box O, Socorro, NM 87801, USA}
\affil{$^{25}$Research School of Astronomy and Astrophysics, Australian National University, Cotter Road, Weston Creek ACT 2611, Australia}
}

\jid{PASA}
\doi{10.1017/pas.\the\year.xxx}
\jyear{\the\year}

\usepackage{aas_macros}
\usepackage{hyperref} 
\hypersetup{colorlinks,citecolor=blue,linkcolor=blue,urlcolor=blue}

\newcommand{\field}[1]{\vspace*{4pt}\noindent {\bf Region #1}}

\hypersetup{draft}

\begin{document}

\begin{frontmatter}
\maketitle

\begin{abstract}
The Variables and Slow Transients Survey (VAST) on the Australian Square Kilometre Array Pathfinder (ASKAP) is designed to detect highly variable and transient radio sources on timescales from 5 seconds to $\sim 5$ years.
In this paper, we present the survey description, observation strategy and initial results from the VAST Phase I Pilot Survey. This pilot survey consists of $\sim 162$ hours of observations conducted at a central frequency of 888~MHz between 2019 August and 2020 August, with a typical rms sensitivity of 0.24~mJy~beam$^{-1}$ and angular resolution of $12-20$ arcseconds.
There are 113 fields, \red{each of which was observed for 12 minutes integration time}, with between 5 and 13 repeats, with cadences between 1 day and 8 months. The total area of the pilot survey footprint is 5\,131 square degrees, covering six distinct regions of the sky. An initial search of two of these regions, totalling 1\,646 square degrees, revealed 28 highly variable and/or transient sources. Seven of these are known pulsars, including the millisecond pulsar J2039--5617.
Another seven are stars, four of which have no previously reported radio detection (SCR~J0533--4257, LEHPM~2-783, UCAC3~89--412162 and 2MASS J22414436--6119311). Of the remaining 14 sources, two are active galactic nuclei, six are associated with galaxies and the other six have no multiwavelength counterparts and are yet to be identified.

\end{abstract}

\begin{keywords}
radio transient sources -- sky surveys -- pulsars -- stars \end{keywords}
\end{frontmatter}

\section{INTRODUCTION }
\label{sec:intro}

Radio variability is produced by a wide range of astronomical sources and phenomena, including the Sun, planets, stellar systems, neutron stars, supernovae, gamma-ray bursts (GRBs) and active galactic nuclei (AGN) \citep{cordes04}. Among the most exciting and anticipated transient sources from extragalactic imaging surveys are neutron star mergers, tidal disruption events, and orphan GRB afterglows \citep{metzger15}.  These can help answer questions regarding the true rates of these events (independent of beaming), their energetics, and the properties of their surrounding media.
However, because the source of the emission is similar in most cases (shock-driven synchrotron emission), distinguishing these sources from other transients can be hard, especially with small numbers of detections well after the sources have faded. Therefore prompt identification and multi-wavelength comparisons are crucial \citep{ofek11,stewart18}.   

At the same time, we must distinguish between extragalactic transients and Galactic foreground sources, primarily low-mass stars and pulsars (analogous to the situation in searches for fast optical transients, where M star flares are the dominant foreground; see e.g., \citealt{cowperthwaite15,ho18,webb20}).  Again, multi-wavelength identification can be crucial, but in some cases the properties of the radio emission itself can be used to identify high brightness temperatures and/or coherent emission that rule out extragalactic sources \citep[e.g.,][]{kaplan08,roy10,pritchard21}.

Early untargeted surveys generally used archival data, and consisted of either many repeats of relatively small sky areas \citep[e.g.,][]{carilli03, bower07}; a few repeats of large sky areas, such as the comparison between the NRAO VLA Sky Survey \citep[NVSS;][]{condon97} and Faint Images of the Radio Sky at Twenty-centimeters \citep[FIRST;][]{becker97} surveys \citep{galyam06}; or inconsistently sampled repeats over a range of sky areas \citep[e.g.,][]{bannister11a}. These surveys typically resulted in a small number of detections, but the historical nature of these detections, and the lack of multi-wavelength data made it difficult to classify or confirm these objects \citep[e.g.,][]{bower07,thyagarajan11}. 

Despite the limitations of these surveys there have been a number of interesting  discoveries. For example, \citet{hyman05} and subsequent papers identified several interesting sources towards the Galactic Centre whose origins are still unknown \citep[see also][]{hyman09}, and \citet{ofek11} used near real-time reduction and multi-wavelength follow-up to identify a single transient that may be consistent with a neutron star merger.  

More recently, the advent of Square Kilometre Array (SKA) pathfinder telescopes and upgrades of established telescopes have allowed a new wave of radio transients surveys \citep[e.g.,][]{driessen20b}.
For example, the Caltech-NRAO Stripe 82 survey \citep[CNSS;][]{mooley16} has used the upgraded Karl G.\ Jansky Very Large Array (VLA) to repeatedly survey an equatorial strip and identified a number of sources, including a tidal disruption event \cite[TDE;][]{anderson20}. 
\citet{stewart16} used repeated observations of the north celestial pole at 60~MHz with the Low Frequency Array (LOFAR; \citealt{vanharlem13}) and found a single transient event of unknown nature. 
The ongoing VLA Sky Survey \citep[VLASS;][]{lacy20} has led to the identification of several sources through comparison with archival surveys \citep{law18,ravi21}.  The ThunderKAT large survey project \citep{fender16} on the MeerKAT telescope \citep{jonas16} will both conduct its own observations and analyse commensal data from other projects, and  has already identified new sources including a chromospherically active K-type sub-giant \citep{driessen20}. Increasingly these detections are being made closer to real-time (within days of observation), aided by improved radio surveys \citep{lacy20,mcconnell20} to allow rapid identification of new (or missing; e.g., \citealt{ravi21}) sources.

This progress at radio wavelengths is coupled with significant improvements (in depth, wavelength coverage, resolution, and temporal sampling) in multi-wavelength coverage, such as the Panoramic Survey Telescope and Rapid Response System (Pan-STARRS; \citealt{chambers16}), the Vista Hemisphere Survey (VHS; \citealt{mcmahon13}), the VISTA Variables in the Via Lactea survey (VVV; \citealt{minniti10}), SkyMapper \citep{keller07}, the Dark Energy Survey \citep[DES;][]{abbott18}, DESI Legacy Imaging Surveys \citep[especially Dark Energy Camera Legacy Survey, or DECaLS;][]{dey19}, the \textit{Wide-field Infrared Survey Explorer} (\wise; \citealt{wright10}) and, especially for stellar science, the \gaia\ mission \citep{gaia16}.  These surveys improve cross-matching and classification for a variety of sources \citep{stewart18}.

The Australian Square Kilometre Array Pathfinder \citep[ASKAP;][]{johnston07,hotan21} is an array of $36 \times 12$-metre prime-focus antennas located at the Murchison Radio-astronomy Observatory in Western Australia. It operates over a frequency range of 0.7--1.8~GHz with an instantaneous bandwidth of 288\,MHz and a \red{nominal} field of view of 31 square degrees. ASKAP was designed as a rapid wide-field survey instrument. The ASKAP Variables and Slow Transients Survey \citep[VAST\footnote{\url{https://vast-survey.org}.};][]{murphy13}
is one of the key survey science projects (SSPs)\footnote{\url{https://www.atnf.csiro.au/projects/askap/ssps.html.}} approved for ASKAP time allocation.

VAST was designed to detect astronomical phenomena that vary on timescales accessible in the ASKAP imaging mode (from $\sim 5$ seconds to several years). 
Since the original survey proposal described by \citet{murphy13}, the scope of the science case has been extended to incorporate detection of radio emission from gravitational wave events (in particular binary neutron star mergers) and fast radio bursts and their host galaxies (in the image domain). More rapidly varying sources (on timescales of less than seconds) are being explored as part of the Commensal Real-Time ASKAP Fast-Transients Survey \citep[CRAFT;][]{macquart10}.

In the Early Science period, we have demonstrated ASKAP's capability to detect time-variable phenomena in imaging mode. We have reported: an initial search for transients and variables in a single 30\,deg$^2$ field using 8 daily epochs and a subset of the final ASKAP array \citep[][also see \citealt{heywood16}]{bhandari18}; the discovery of a new millisecond pulsar J1431$-$6328 \citep{kaplan19}; a sample of 23 previously undetected flaring stars \citep{pritchard21}; and radio bursts from UV Ceti \citep{zic19} and Proxima Centauri \citep{zic20}. In addition, ASKAP was used to follow-up several gravitational wave events from LIGO/Virgo, notably GW190814 \citep{dobie19,abbott20a}.

During 2019--2020, each of the SSPs was allocated 100 hours of observing time to conduct a Phase I Pilot Survey with ASKAP. In 2021, another 100~hours have been allocated to conduct a Phase II Pilot Survey. The purpose of these pilot surveys is to test the observing strategy and implementation, data processing capabilities, and scientific analysis in preparation for the full surveys, which are expected to begin in 2022. 

In this paper we describe the VAST Phase I Pilot Survey, present some initial results, and discuss the plan for the VAST Phase II Pilot Survey. In Section~\ref{s_obs} we discuss our observational specifications and scheduling; in Section~\ref{s_pipeline} we discuss our transient detection pipeline software; in Section~\ref{s_search} we present results from an untargeted variability search on two of the VAST Phase I Pilot Survey regions (3 and 4); and in Section~\ref{s_summary} we discuss our future plans.

\section{OBSERVATIONS AND SURVEY STRATEGY}\label{s_obs}

\subsection{The Rapid ASKAP Continuum Survey}
In addition to our dedicated VAST  observations, we incorporated data from the Rapid ASKAP Continuum Survey \citep[RACS;][]{mcconnell20} as our first epoch. RACS is the first large-area survey to be conducted with the full 36-antenna ASKAP telescope. 
When complete, RACS will cover the sky south of declination $+51^\circ$ (a total sky area of 36\,656 deg$^2$), across three ASKAP bands (centred on 888, 1296 and 1656\,MHz). The angular resolution is $15-25$\,arcsec in the low band and $8-$27\,arcsec in the mid band, depending on the declination. 

The first data release of RACS consists of 903 images covering the sky south of declination $+41^\circ$, observed in the low band (centred on 888\,MHz). Each field was observed for $\sim 15$\,min, achieving a typical rms noise of 0.25\,mJy\,beam$^{-1}$. All four instrumental polarisation products (XX, XY, YX, and YY) were
recorded to allow images to be made in Stokes parameters I, Q,
U, and V. The data were processed using the \askapsoft\ pipeline \citep{cornwell11} and are available through the CSIRO ASKAP Science Data Archive \citep[\casda\footnote{\url{https://research.csiro.au/casda/}.};][]{chapman17} under project code AS110. 

We used the RACS low-band survey as `epoch 0' of the VAST Phase I Pilot Survey (VAST-P1) which was conducted entirely in the lowest frequency band. We will use the RACS mid-band survey as `epoch 0' of the VAST Phase II Pilot Survey (VAST-P2), which will be conducted in both the low and mid-bands. This allows us to take advantage of the RACS tiling \red{pattern} and survey design, which had very similar parameters to those required for VAST. An important note is that the publicly-released RACS images have had an additional flux correction applied, as described by \citet{mcconnell20}, so any epoch 0 flux densities reported in this work may differ from the (improved) published RACS catalogue (Hale et al., submitted).

\red{In in this paper, we use the following terminology to describe the different components of our survey and others:
\begin{description}
\item[beam footprint] The arrangement of phased-array feed (PAF) beams used to cover a single field.  We use the \texttt{square\_6x6} and \texttt{closepack36}   beam footprints (see \citealt{hotan21}).
\item[field] A single observation or pointing.
\item[tiling pattern] How individual fields are placed to cover a larger area.
\item[region] A grouping of pointings (usually but not always covering  adjacent) meant to cover a particular area of the sky or type of source.
\item[survey footprint] The total sky area covered by a survey (for instance, DES) or one epoch of a survey (as in VAST-P1).
\end{description}}

\subsection{Phase I Pilot Survey observations}

\begin{figure*}
	\includegraphics[width=\textwidth,clip]{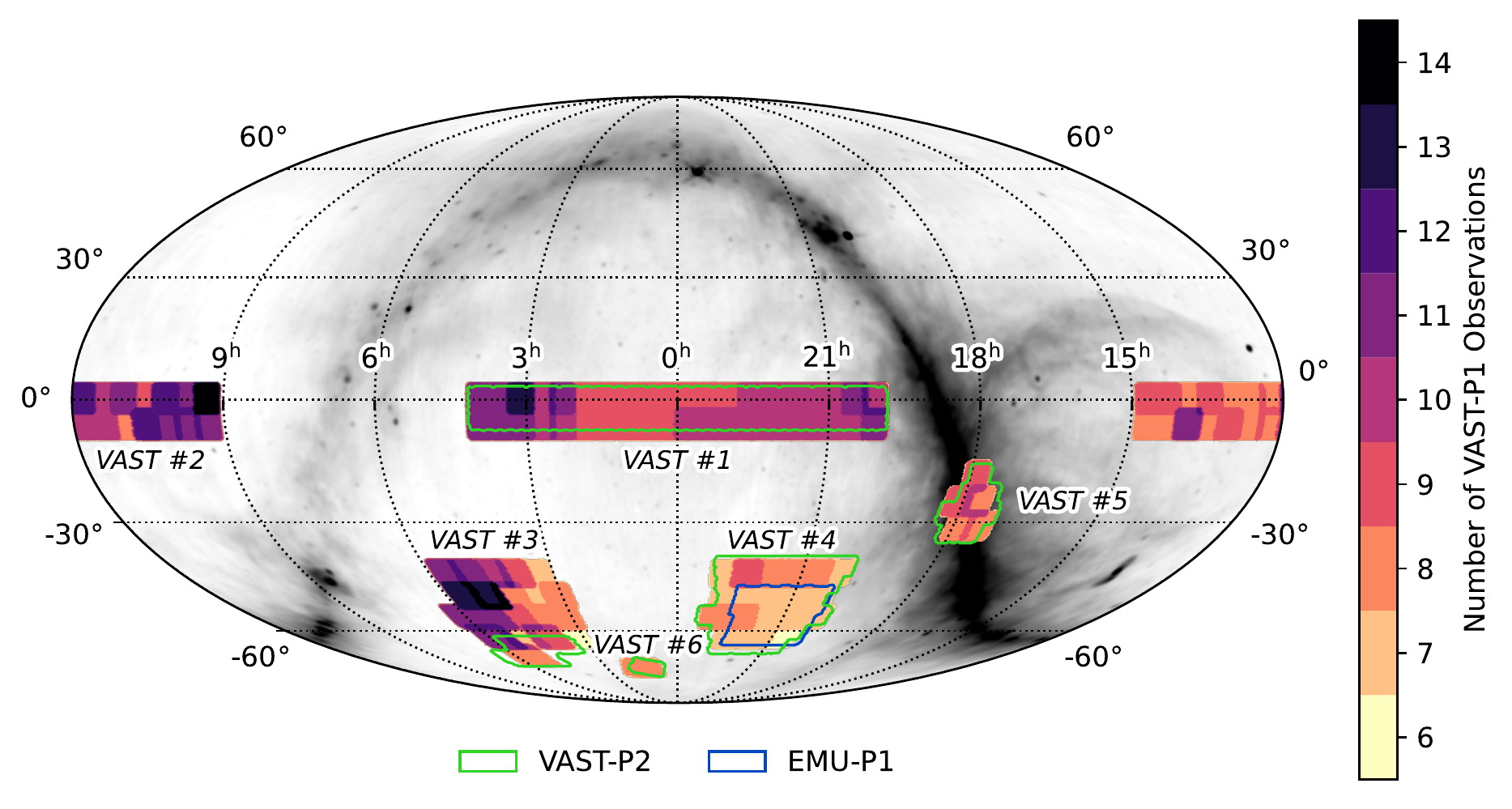}
    \caption{The VAST-P1 \red{survey} footprint, showing the number of observations of each field. The green region shows the planned   \red{survey} footprint of the VAST-P2 mid-band observations. \red{VAST-P2 low-band will cover the same \red{survey} footprint as VAST-P1}. The sky map is plotted with J2000 equatorial coordinates in the Mollweide projection and the background diffuse Galactic emission at 887.5\,MHz is modelled from \citet{zheng17} using \citet{price16}.}
    \label{f_sky}
\end{figure*}

VAST-P1 consists of 113 fields taken from the RACS low-band \red{tiling pattern}. The survey coverage of VAST-P1 is shown in Figure~\ref{f_sky}. The survey consists of six regions, chosen to test some of the different VAST science goals. These are summarised below.

\field{1:} An extragalactic region of 40 fields  (1840\,deg$^2$) along an equatorial strip centred at RA of $0^{\rm h}$. This was chosen to overlap with the CNSS, which itself had considerable overlap with FIRST.  In addition, this overlaps with part of the DES/DECaLS coverage.

\field{2:} A second extragalactic region of 28 fields  (1300\,deg$^2$) along an equatorial strip centred at RA of $12^{\rm h}$. The equatorial regions have the advantage of greater multi-wavelength coverage from telescopes in the northern hemisphere.   This region in particular has good overlap with FIRST and DECaLS.

\field{3:}
An extragalactic region of 19 fields  (820\,deg$^2$) chosen within the larger DES \red{survey} footprint, centred at ${\rm RA}=4.5$\,hr, ${\rm Dec}=-50\deg$.

\field{4:} A second region of 19 fields  (830\,deg$^2$) within the DES \red{survey} footprint, centred at ${\rm RA}=22$\,hr, ${\rm Dec}=-50\deg$. Region 4 also overlaps with the Evolutionary Map of the Universe \citep[EMU;][]{norris11} Phase I Pilot Survey and the Polarisation Sky Survey of the Universe's Magnetism \citep[POSSUM;][]{gaensler10} Phase I Pilot Survey.

\field{5:} A small 5-field  (265\,deg$^2$)  region covering the Galactic Centre.

\field{6:} A small 2-field  (135\,deg$^2$) region covering the Small and Large Magellanic Clouds.

\vspace{4pt}
\red{Note that the region area above is not the number of fields times 31\,deg$^2$ for two reasons.  First, the instantaneous field-of-view is typically more than 31\,deg$^2$ as it includes sky sampled by the PAF beams which is outside the region over which the PAF can point (which is the origin of the 31\,deg$^2$ number).  The field-of-view is frequency dependent\footnote{See \url{https://confluence.csiro.au/display/askapsst/ASKAP+Observation+Guide}, Figure~7.4.} but can be up to 64\,deg$^2$ at the low-band frequency, including lower-sensitivity wings of the PAF beams.  Second, when fields are adjoining the lower-sensitivity wings overlap, improving sensitivity but reducing total sky coverage.}

The ASKAP observing specifications in the context of VAST-P1 are summarised in Table~\ref{t_specs}, and 
the full set of VAST-P1 observations is summarised in Table~\ref{t_obs}.
Our survey strategy was to observe five epochs of the VAST-P1 \red{survey} footprint, with each epoch separated by months, and one pair of epochs separated by a single day. However, there were also extra test observations conducted, and some epochs in which extra fields were observed to accommodate observations where the schedule initialisation failed (or was excessively delayed). Hence, in addition to the \red{six} full epochs (epochs 1, 2, 8, 9, \red{12 and 13} in Table~\ref{t_obs}), there are a number of epochs that have partial coverage of the VAST-P1 \red{survey} footprint. These epochs are labelled with an `x' in their name in Table~\ref{t_obs} (3x, 4x, 5x, 6x, 7x, 10x and 11x). \red{The data for epoch 12 (marked with as asterisk in Table~\ref{t_obs}) only became available after submission of this paper. We have included it in the table for completeness but the data is not included in any of the analysis presented in this paper.}

\begin{table*}
\caption{VAST observing parameters. Note the number of epochs for VAST-P2 (as well as the minimum and maximum spacing) are planned estimates, and may change when the survey is conducted. The image rms and total area for VAST-P2 (mid) are estimated from early RACS-mid observations; these estimated values are marked in italics. \red{See \citet{hotan21} for details about the beam footprints.}}
\centering
\begin{tabular}{@{}lccc@{}}
\hline
Parameter & VAST-P1 & VAST-P2 (low) & VAST-P2 (mid) \\
\hline
Centre Frequency & 888 MHz & 888 MHz & 1296 MHz \\
Bandwidth & 288 MHz & 288 MHz & 288 MHz \\
Integration per tile & 12 min & 12 min & 12 min \\
\red{Beam} footprint & square\_6x6 & square\_6x6 & closepack36 \\
Beam spacing & $1.05^\circ$ & $1.05^\circ$ & $0.9^\circ$ \\
Total \red{survey} footprint area & 5131 deg$^2$ & 5131 deg$^2$ & {\it 2638 deg$^2$} \\
Number of epochs & 5--13 & {\it 2} & {\it 3 } \\
Minimum spacing & 1\,day & {\it 1 month } & {\it 2 months } \\
Maximum spacing & 12\,months & {\it 1\,month} & {\it 4\,months } \\
Image rms per epoch & 0.24\,mJy\,beam$^{-1}$ & 0.24\,mJy\,beam$^{-1}$ & {\it $\mathit{\lesssim 0.2}$\,mJy\,beam$^{-1}$}\\
\hline
\end{tabular}
\label{t_specs}
\end{table*}

% Total of 813 fields excluding RACS
% Total time = 162 hours 
The total combined area of a full VAST-P1 epoch is 5131\,deg$^2$. Each field was observed for 12~minutes integration time (rather than 15~minutes for RACS). Excluding RACS (as epoch 0) we observed 813 fields, for a total observing time of $\sim162$\,hours. In this paper we present results from two of the regions (3 and 4). The rest of the VAST-P1 \red{survey} footprint will be analysed in subsequent papers.

\begin{table}
\caption{Summary of VAST-P1 observations, giving the number of fields in each epoch, the start and end dates for the epoch, and the total sky area. Epochs with an `x' in the name only have partial sky coverage, as discussed in the text. Epoch 0 is the RACS survey. \red{Epoch $12^*$ was only available after submission of this paper and is included here for completeness but is not included in any of the analysis.}}
\label{t_obs}
\begin{tabular}{crllr}
\hline
Epoch &  Num. &   Start Date &   End Date & Area \\
      &  Fields     &     & & deg$^{2}$ \\
\hline
0   &        113 & 2019 Apr 25 & 2019 May 08 & 5131 \\ 
1    &        113 & 2019 Aug 27 & 2019 Aug 28 & 5131 \\
2    &        108 & 2019 Oct 28 & 2019 Oct 31 & 4905 \\
3x   &         43 & 2019 Oct 29 & 2019 Oct 29 & 2168 \\
4x   &         34 & 2019 Dec 19 & 2019 Dec 19 & 1672 \\
5x   &         81 & 2020 Jan 10 & 2020 Jan 11 & 3818 \\
6x   &         49 & 2020 Jan 11 & 2020 Jan 12 & 2400 \\
7x   &         33 & 2020 Jan 16 & 2020 Jan 16 & 1666 \\
8    &        112 & 2020 Jan 11 & 2020 Feb 01 & 5097 \\
9    &        112 & 2020 Jan 12 & 2020 Feb 02 & 5097 \\
10x   &         13 & 2020 Jan 17 & 2020 Feb 01 & 803 \\
11x   &         11 & 2020 Jan 18 & 2020 Feb 02 & 695 \\
\red{$12^*$}    &       109 & 2020 Jun 19 & 2020 Jun 21 & \red{5100} \\
\red{13}    &       104 & 2020 Aug 28 & 2020 Aug 30 & 5028 \\ 
\hline
\end{tabular}
\end{table}

\subsection{Phase II Pilot Survey strategy}
The VAST Phase II Pilot Survey will commence in 2021. The aims of VAST-P2 are: to build on the scientific analysis from VAST-P1; to test faster turn-around of imaging and transient detection; and to demonstrate commensality with ASKAP mid-band surveys by observing some epochs at 1296\,MHz. 

We plan to observe five epochs spread over the six months of the pilot program. We will observe two epochs at low-band, using the same \red{survey} footprint as VAST-P1 and the {\tt square\_6x6} beam footprint. The remaining three epochs will be observed in the ASKAP mid-band. This will use the \red{tiling pattern} from the RACS mid-band survey and the {\tt closepack36} beam \red{footprint \citep{hotan21}}. 

The size of the primary beams scales proportionally with wavelength. Since we tile the sky with \red{beam} footprints that provide nearly uniform sensitivity across the observed area, our mid-band survey will require a different tiling \red{pattern} where each \red{beam} footprint will cover less area than the low-band \red{beam} footprints. Just as we have done for our low-band survey, we select our mid-band \red{tiling pattern} from the RACS mid-band survey tiling \red{pattern}. Note that since the mid-band \red{beam} footprints are smaller, the low- and mid-band \red{field} centres are not aligned.

Subtracting the duration of two low-band epochs from the total VAST-P2 observing time will leave enough time to observe 3 epochs of 91 mid-band fields. This will not be enough to cover the entire low-band survey area, so we have iteratively selected mid-band fields from RACS to optimise the overlap with the low-band fields from the following VAST regions in order: 5 (Galactic Centre), 6 (Magellanic Clouds), 1 (equatorial strip), then 4 (DES/EMU/POSSUM). This results in 88 mid-band fields. The remaining 3 mid-band fields have been selected from region 3 (DES) as it is adjacent to the already selected region 6 fields that cover the LMC. The full set of selected mid-band fields is shown in Figure~\ref{f_sky}.

\subsection{Data products}
\begin{figure*}
    \includegraphics[width=\textwidth]{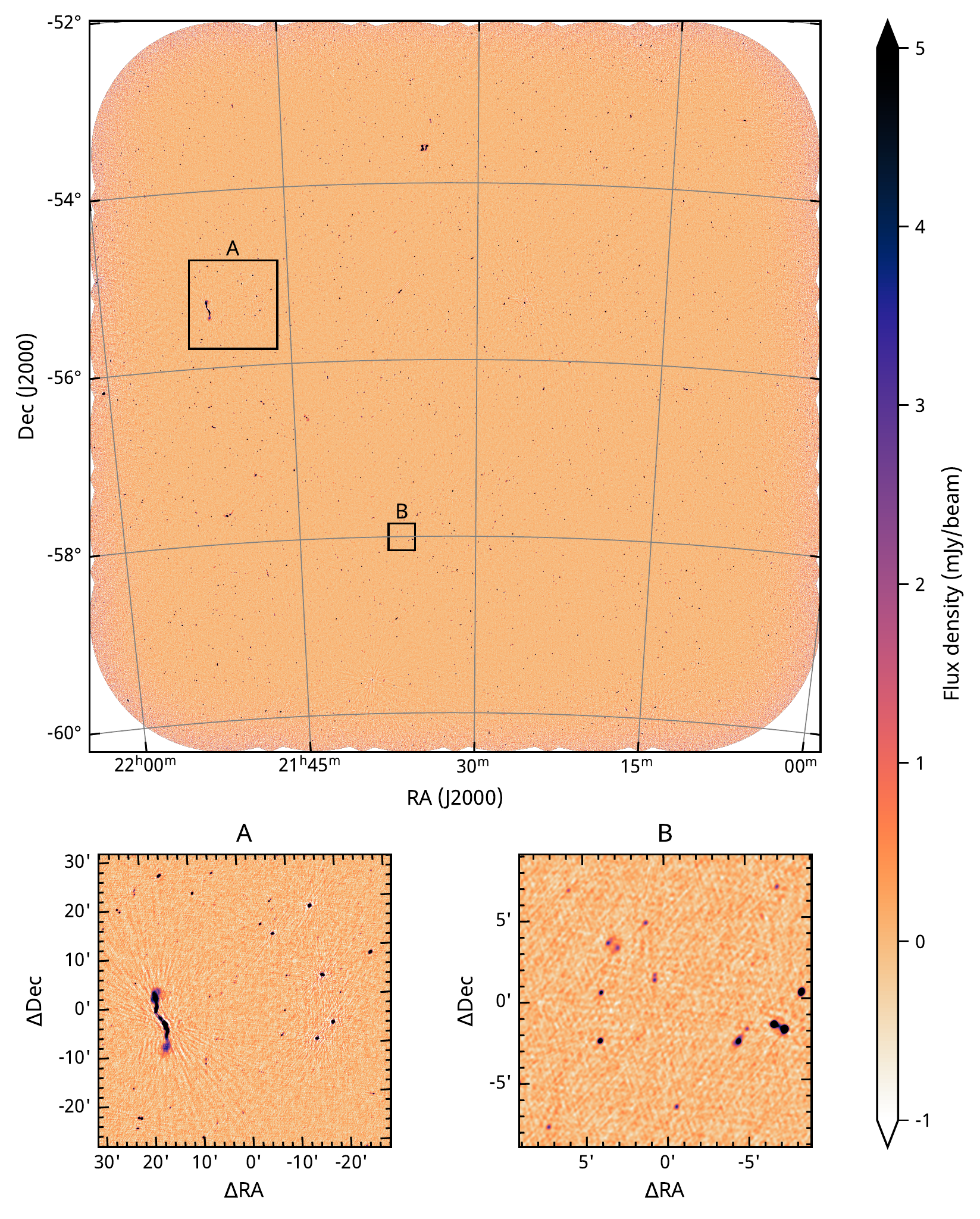}
    \caption{A VAST image within Region 4 (field VAST\_2131$-$56A) from epoch 12 with two cutouts. \textit{Cutout A}: a $1\deg$ image centered on (J2000) $\alpha$ = 21:49:26.5, $\delta$ = --55:17:52.87 containing several bright sources, including the large radio galaxy 2MASX~J21512991--5520124. \textit{Cutout B}: a $0.3\deg$ image centered on (J2000) $\alpha$ = 21:36:18.9, $\delta$ = --58:00:12.68 containing a range of source morphologies.} \label{f_image}
\end{figure*}

VAST-P1 data was processed using the \red{same calibration and imaging strategy as that used by the RACS survey \citep{mcconnell20}, the only difference was that VAST used a standard Gaussian primary beam correction whereas RACS applied a position-dependent correction based on holography data to correct for slight non-Gaussianity of the PAF beams. The use of holographic data to correct for primary beam attenuation was made available for general processing in the ASKAPsoft pipeline after most VAST processing for VAST-P1 was completed, however, it was used in the processing of epoch 12.} There are two main image data products:
\begin{enumerate}
    \item The individual field (or `tile') images produced by \askapsoft. These were created by combining the 36 individual beam images. We produced multi-frequency synthesis (MFS) images for all four Stokes parameters I, Q, U and V. Two Taylor-terms are used in Stokes I imaging to consider the spectral \red{slope} of field sources across the observed band.
    \item Combined images made by mosaicing the individual field images together (these are not produced by \askapsoft) across each region of VAST-P1. This means that images at different times {\it within} a given epoch are combined,  so they can only be used for epoch-to-epoch comparisons.  
\end{enumerate}
A typical VAST-P1 image is shown in Figure~\ref{f_image}. When imaging, we removed short ($< 100$~m) baselines in order to; (i) to minimise solar interference; and (ii) as there is insufficient uv-coverage to reliably recover extended structure in short VAST-P1 observations.

\begin{figure}
    \centering
	\includegraphics[width=\columnwidth]{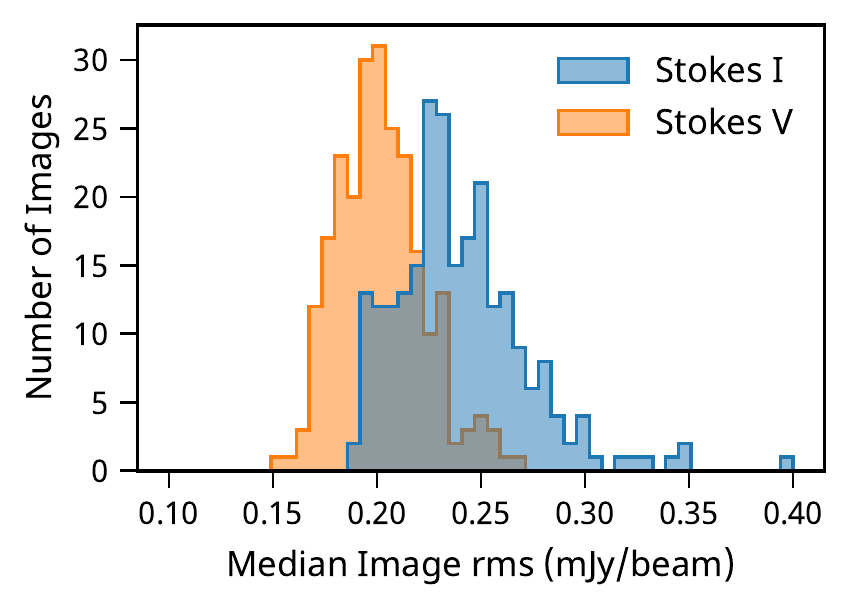}
    \caption{Distribution of median image rms values (computed over the central half of each image) for each field in each epoch of regions 3 and 4.  We plot the rms values from Stokes I (blue) and Stokes V (orange).}
    \label{f_rms}
\end{figure}

A histogram of the rms noise in each individual VAST image in regions 3 and 4 is shown in Figure~\ref{f_rms}, where we separately show Stokes I and Stokes V.  The latter is dominated by thermal noise in the images, while the former also has significantly more contributions from source confusion, sidelobe confusion, and deconvolution artifacts.  The median rms is $0.24\,{\rm mJy\,beam}^{-1}$ in Stokes~I, while it is $0.20\,{\rm mJy\,beam}^{-1}$ in Stokes~V.

In addition to images, the \askapsoft\ pipeline also produces a catalogue of groups of pixels of contiguous emission (`islands'), a catalogue of Gaussian components fit to the islands (`components', where each island may be fit with one or more components), and an rms map and a median map corresponding to each field. These are created by the \selavy\ source finding software \citep{whiting12} incorporated into \askapsoft. We ran \selavy\ separately on our combined images, and used these additional data products as inputs to our VAST transient detection pipeline, as described in Section~\ref{s_pipeline}. We used the default parameter settings in \askapsoft, so sources were considered `detected' if they exceeded 5 times the local rms (roughly 1.2\,mJy in Stokes I for an unresolved source).

Two of the VAST-P1 epochs (8 and 9) were observed on consecutive days, and were matched in local sidereal time (LST). This was done to explore whether LST-matching improved the ability to search for transients using image subtraction. The observations for the matched epochs were taken over a period of two weeks, but within that period each individual field was LST matched and observed on consecutive days.

\subsection{Data quality analysis}
Before images were released in \casda, we performed some quality control checks of the astrometric accuracy and flux density scale (both relative and absolute). In addition, we visually inspected each image to check the overall data quality (on the basis of this, a single field was excluded: VAST\_0534--43A in epoch 7x, which contains Pictor A). In this section, we summarise the results for the two VAST-P1 regions (3 and 4) presented in this paper (see Section~\ref{s_results}).

\subsubsection{Astrometric accuracy}\label{s_astrometry}
To evaluate the astrometric accuracy of our images, we extracted bright (SNR $>=7$), compact\footnote{Compactness is determined following the definition by (Hale et al., submitted) of an integrated to peak flux ratio of $S_I/S_P < 1.024 + 0.69\times \mathrm{SNR}^{-0.62}$.} sources using \selavy\ that are isolated by a minimum of $150$ arcsec from other sources, and then crossmatched them with the International Celestial Reference Frame (ICRF) catalogue \citep{charlot20}. The ICRF consists of 4536 radio
sources with accurate positions measured using very long baseline interferometry (VLBI). 

\begin{figure*}
	\includegraphics[width=\linewidth]{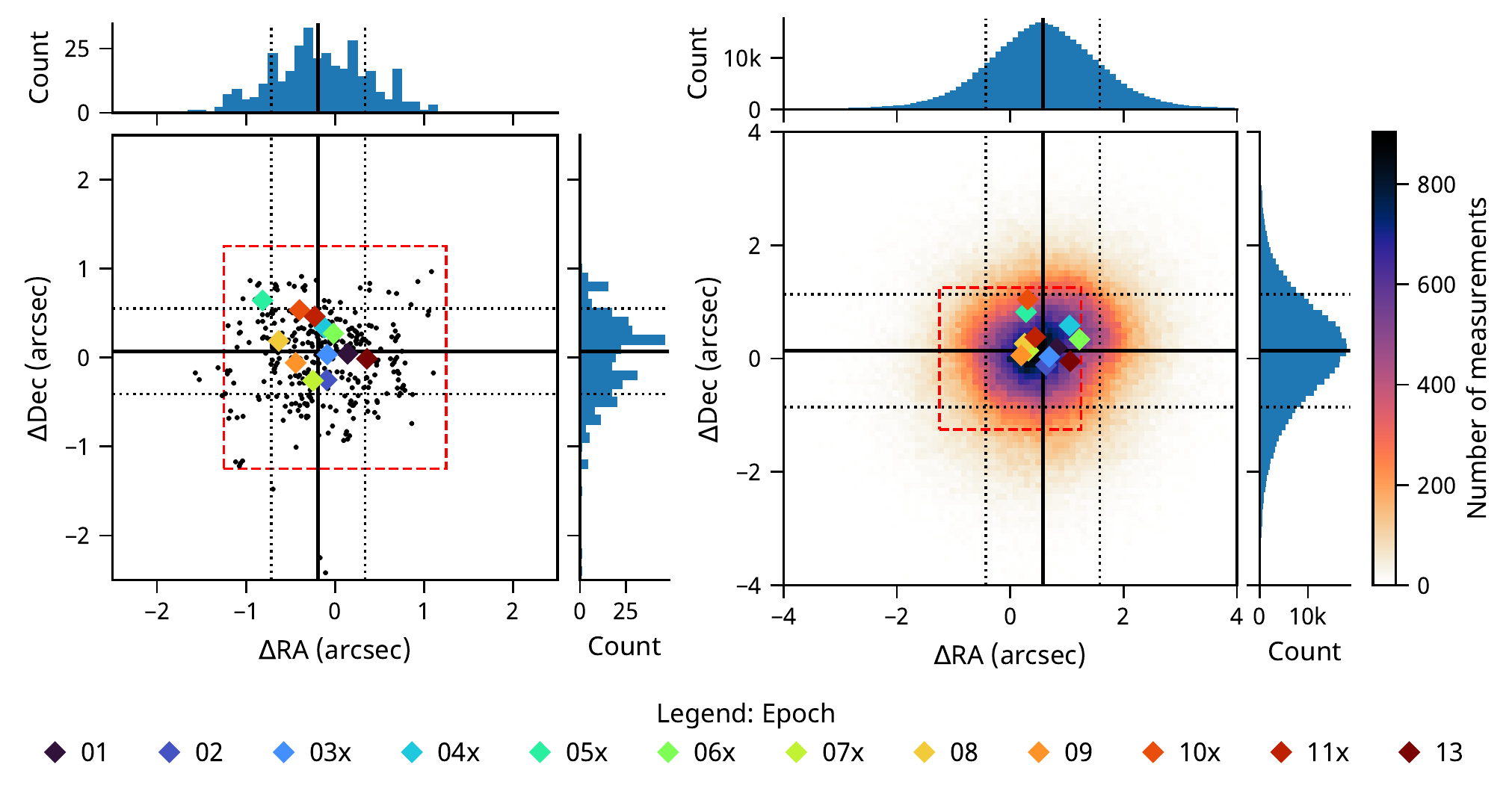}
    \caption{Astrometric accuracy for compact sources in regions 3 and 4 of VAST-P1 compared to {\it Left}: sources from the ICRF catalogue and {\it Right}: sources from the RACS catalogue, as described in Section~\ref{s_astrometry}. The image pixel size of $2.5\times 2.5$ arcsec is shown as a red dashed box. The median offset for each individual epoch is shown with coloured markers. \red{The solid lines show the overall median offsets, and the dashed lines are the median $\pm 1$ standard deviation.}  For the comparison with RACS, the background is a 2D histogram of the source counts where the colour scale represents the number of sources per bin. }
    \label{f_positions}
\end{figure*}

The left panel of Figure~\ref{f_positions} shows the offsets for VAST-P1 sources associated with the 41 ICRF sources in regions 3 and 4 (a total of 331 comparisons across all epochs). The median and standard deviation of the positional offsets is $-0.19 \pm 0.53$\,arcsec in right ascension and $0.07 \pm 0.48$\,arcsec in declination, \red{with a standard error of $29$\,milliarcsec in both coordinates}.
The median positions of each epoch of VAST-P1 are shown as coloured markers. 

We also compared the VAST-P1 sources with the positions in the published RACS catalogue (Hale et al., submitted). There were $378\,823$ unique measurements of $32\,769$ compact, isolated RACS sources across all epochs within regions 3 and 4 with SNR $>=7$. The results of this comparison are shown in the right panel of Figure~\ref{f_positions}. The median and standard deviation of the positional offsets are $0.58 \pm 1.01$\,arcsec in right ascension and $0.14 \pm 0.99$\,arcsec in declination, \red{with a standard error of $1.6$\,milliarcsec in both coordinates}.

\subsubsection{Flux density scale}\label{s_fluxes}
\begin{figure}
	\includegraphics[width=\columnwidth]{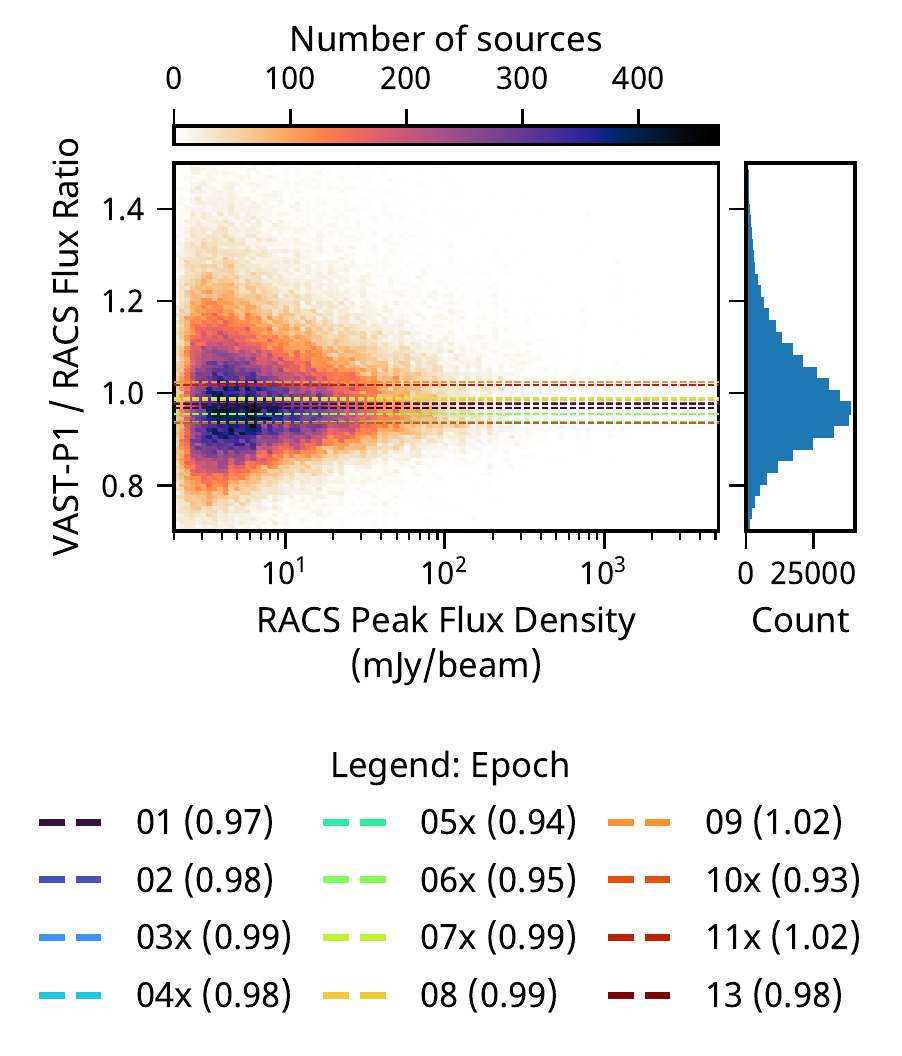}
\caption{Absolute and relative flux density scale comparison for $378\,823$ sources in regions 3 and 4, comparing those measured in each VAST-P1 epoch to RACS (as discussed in Section~\ref{s_fluxes}). The median ratio for each epoch is shown by the dotted lines. The background is a 2D histogram of the source counts where the colour scale represents the number of sources per bin. The overall median ratio is 0.98 with a scatter of 0.15.
}
    \label{f_fluxes}
\end{figure}

In addition to the absolute flux density scale, for variability analysis it is important that the flux density scale from epoch to epoch is consistent. To evaluate both of these, we crossmatched compact, isolated sources with $\mathrm{SNR}>=7$ in each epoch of VAST-P1 with sources in the published RACS catalogue. Figure~\ref{f_fluxes} shows the flux density ratio for these $378\,823$ sources.

Nine of the twelve epochs have a median VAST-P1/RACS flux density ratio between 0.97 and 1.03 (within $3\%$). Epoch 10x has a ratio of 0.93 and Epochs 05x and 06x have ratios of 0.94 and 0.95 respectively. The overall ratio is 0.98 with a $1\sigma$ scatter of 0.15. This scatter is probably the result of a few differences in the way that RACS and VAST-P1 data were processed. The most significant of these is that the published RACS images had a holography correction applied that improved the flux density scale, particularly in the outer regions of images. 
These discrepancies will be corrected in future analyses, but for the current work our conservative variability thresholds (see below) mean that they are not significant.

\subsubsection{Known issues}
As discussed above, the observations presented here are part of a pilot study for a full survey.  Therefore some of the observations and analysis used were less than optimal, and evolved as we discovered deficiencies.  These lessons will inform the full VAST survey.  Nonetheless, the data are useful for scientific analysis, but we note the various issues below.

From epoch 4x onwards, we made all observations within $\pm 1$\,h of meridian transit to ensure a consistent synthesised beam size for all fields. In epochs 1--3x, this did not occur and some fields were observed at large zenith angles, resulting in very extended synthesised beams.

From epoch 5x onwards, an intermittent system issue at the initialisation of a new field observation caused random phase delays on a few antennas on a small subset of field observations. These were identified after processing, but could not be corrected in all cases. Furthermore, even for corrected fields, this occasionally resulted in increased astrometric and flux scale errors. Fields that were affected by this issue are highlighted in the release notes in \casda. Other more subtle changes in the bandpass and beam-shape may also have occurred in early observations as beam-forming and use of the on-dish calibrator improved over the course of the pilot survey. As a result, more recent data tend to be of higher quality than earlier data.

The process used to obtain an initial deconvolution model for use in self-calibration in the ASKAP pipeline can introduce a slight positional offset in each beam image. This is because self-calibration will tend to snap to a dominant field source and place it at a pixel centre. This not only causes astrometric errors, up to half a pixel (1.25\,arcsec) in extent, but can also cause apparent extension of a source if the shift is in opposite directions for a source in two overlapping beams.

The source finding for the VAST pipeline input was performed on final mosaiced images. 
In early pilot survey images, it was assumed that the point spread function (PSF) of each beam would be consistent in each field. In practice, the PSF can vary from beam to beam as a result of differences in flagging and elevation. As only a single PSF is recorded in the image header, typically the PSF associated with beam 0, this can affect the accuracy of the flux scale across the entire mosaic.

Also, note that the RACS flux density scale correction (as described in Appendix A of \citealt{mcconnell20}) has not been applied to the VAST data. Hence we also used the pre--flux-corrected RACS data in epoch 0, for consistency. For future work, we plan to convolve all fields to a consistent PSF and apply these corrections to ensure consistency between RACS-low, RACS-mid and both the VAST pilot surveys.

\subsection{Data availability}
The raw data and individual field images for the entire VAST-P1 survey are publicly available through \casda\ under project code AS110 for RACS and AS107 for VAST. Epoch 4x of VAST-P1 was conducted as a test observation and is available under project code AS113. 
Higher order data products such as variability catalogues will be released once full analysis is complete.

\section{TRANSIENT DETECTION PIPELINE}\label{s_pipeline}
The primary technical goal of the VAST Pilot Surveys is to demonstrate the capability to detect highly variable and transient sources. One of the main open-source software packages for image-based radio transient detection is the LOFAR Transients Pipeline \citep[\trap;][]{swinbank15}. \trap\ can be applied to radio image data from any facility, and has been used successfully in a number of transient projects \citep[e.g.,][]{stewart16,driessen20,sarbadhicary20}. 

Testing \trap\ on the early ASKAP datasets revealed some scalability issues. After investigating a number of options, we decided to implement a new pipeline to search ASKAP data in an efficient way. Most of the key logic of this pipeline follows the approach of \trap, but with some implementation differences that significantly improve the performance.

The main VAST Pipeline technology stack is: Python 3.7+, \software{PostgreSQL}\footnote{\url{https://www.postgresql.org}.}~$12+$, \software{Astropy}\footnote{\url{https://www.astropy.org}.}~$4+$, \software{Django}\footnote{\url{https://www.djangoproject.com}.}~$3+$,
\software{Pandas}\footnote{\url{https://pandas.pydata.org}.}~$1.2+$,
\software{Dask}\footnote{\url{https://dask.org}.}~$2+$ and \software{Bootstrap}\footnote{\url{https://getbootstrap.com/}.}~4.
A more complete technical description of the pipeline implementation is given by \citet{pintaldi21}. In this section we discuss the scientific functionality and resulting data products in the context of the results presented here.

\subsection{Pipeline processing method}
The VAST pipeline takes as input a set of images, rms maps and mean (background) maps (all as FITS files) from \selavy\ as well as the \selavy\ source finding component list. These data products are all default outputs from the \askapsoft\ continuum imaging process, which means the VAST pipeline can be run immediately after the \askapsoft\ data reduction is completed.

For the work presented in this paper, we used the combined images (created by mosaicing individual fields together) as the input. This means that only epoch-to-epoch timescales could be explored. The entire \selavy\ source list was read in, but any duplicate sources from overlapping fields within the same epoch were filtered out using a tight crossmatching radius of 2.5\,arcsec. In principle it would also be possible to search for variability within epochs in the regions where fields overlap: this will be the subject of future analysis. Investigating faster variability (within a single observation, for example, \citealt{wang21}) will also be explored in future work.

We ran the VAST pipeline on these combined images, following the steps below:
\begin{enumerate}
    \item {\bf Image ingest:} Image metadata and \selavy\ catalogues were ingested for each image.
    \item {\bf Uncertainties} on flux densities and positions for every source component were calculated following \citet{condon97}.\footnote{The uncertainties calculated by \selavy\ were not correct at the time we ran the ASKAPSoft pipeline. This is being addressed for a future release (Whiting, private communication).}
    \item {\bf Source association} was done by crossmatching each epoch to the next epoch using the de Ruiter method (\citet{scheers11}, based on \citealt{deruiter77}). We used the default beamwidth limit of 1.5 and the search radius of 5.68 (these are both unitless values). The beam size that is used for the epoch mode association should be equal to the largest beam contained in the epoch images. % (essentially there is a hard cut off for association of 1.5x beam size).
    \item {\bf Forced extractions} (where a flux density is measured at a specified position) were performed for sources that were not detected in every epoch\footnote{\url{https://github.com/askap-vast/forced_phot.}}, so that a complete light curve could be built for each source. Where there were multiple observations of a given source within an epoch (e.g., due to image overlap), the forced extractions were taken from the image whose centre was nearest to the source position.
    \item {\bf New sources} that had not appeared in previous epochs were identified. 
    \item {\bf Variability metrics} were calculated for all sources identified by the pipeline (see Section~\ref{s_var}). The forced extraction measurements were incorporated when calculating all variability metrics.
    \item {\bf Results} were saved in a database and output to \software{parquet}\footnote{\url{https://parquet.apache.org/documentation/latest/}.}  files, an open source columnar data format that allows flexible custom post-processing \citep{vohra16}.
\end{enumerate}
The resulting data products are a light curve for every source detected in one or more epochs, and a set of standard variability metrics. Figure~\ref{f_lightcurves} shows light curves for the highly variable sources presented in this paper.

The pipeline results are available to users via either an interactive web application (including some basic multi-wavelength information), or through the \software{parquet} files that can be read by a variety of data-analysis software packages. The \vasttools\ \software{python} package, described in Section~\ref{s_tools}, offers a streamlined method to explore the pipeline results in a \software{Jupyter} notebook environment. 
For more detailed information about the VAST pipeline, see the publicly available software and documentation\footnote{\url{https://github.com/askap-vast/vast-pipeline}.}.

\subsection{Variability metrics}\label{s_var}
The VAST pipeline identifies highly variable and transient sources using the two key light curve parameters identified by \citet{rowlinson19}. 
These are: the modulation index $V$, a measure of the variability in a light curve; and the reduced $\chi^2$ relative to a constant model $\eta$, a measure of the significance of the variability. These parameters are defined as:

\begin{equation}\label{e_var2}
V = \frac{1}{\overline{S}}\sqrt{\frac{N}{N-1} (\overline{S^2} - \overline{S}^2)},    
\end{equation}

\begin{equation}\label{e_var1}
\eta = \frac{N}{N-1}\left( \overline{w S^2} - \frac{\overline{w S}^2}{\overline{w}}\right),
\end{equation}
where $N$ is the number of data points, the flux density and uncertainty on the $i$-th epoch are $S_i$ and $\sigma_i$, means are denoted by overbars (i.e., $\overline{S} \equiv \frac{1}{N}\sum_i S_i$) and in calculating $\eta$ the flux density measurements are weighted by the uncertainties according to $w_i=1/\sigma_i^2$ (also see \citealt{swinbank15}).

The parameter $V$  selects sources with high variability without assessing the statistical significance of that variability. In contrast, the parameter $\eta$ acts as a measure of the statistical significance of that variability without knowledge of its absolute amplitude. Used in combination, these parameters select a sample of statistically significant, highly variable objects.

The pipeline also allows for two-epoch variability searches (where the variability is calculated based on pairs of observations), but this functionality was not used in the results presented here.

%https://arxiv.org/pdf/1808.07781.pdf

\subsection{VAST Tools software}\label{s_tools}
\vasttools\footnote{\url{https://github.com/askap-vast/vast-tools}.} is a \software{python} package containing tools for accessing and exploring VAST data products. The current functionality at the time of latest release (Version 2.0.0) is outlined below. \vasttools\ can be used via a command line interface or in \software{Jupyter} notebooks. \vasttools\ can also be used as an interface to analyse results from a VAST Pipeline run. 

\subsubsection{Survey information}
The \vasttools\ package includes the complete details of the spatial and temporal coverage of VAST-P1. The beam centres for all fields are listed in comma-separated-values (CSV) files on a per-epoch basis, while Hierarchical Equal Area isoLatitude Pixelisation (HEALPix; \citealt{gorski99}) Multi-Order Coverage (MOC\footnote{\url{http://ivoa.net/documents/MOC/}.}; \citealt{fernique19}) maps contain the spatial coverage of each survey field, and each epoch. We have also created a space-time MOC for the full pilot survey coverage. The MOCs can be accessed though a \software{mocpy}\footnote{\url{https://github.com/cds-astro/mocpy/}.} wrapper included in the \vasttools\ package, which also allows users to load MOCs and also query VizieR tables for sources within the MOC coverage.

\subsubsection{Querying coordinates}
The Query class allows users to search the VAST pilot for observations of a given location.
Three main query types exist: (1) finding the fields containing the specified coordinates; (2) searching for radio emission at the specified coordinates; and (3) finding all sources within a given radius of the specified coordinates. Query types 2 and 3 require VAST tools to be run on a machine with access to the relevant FITS files and \selavy\ catalogues.

VAST tools can crossmatch coordinates with the \selavy\ catalogues, and produce light curves, FITS cutouts and postage stamp images with various overlays (e.g., Figure~\ref{f_cutouts}). Forced fitting is available, and there are options to produce light curves with upper limits (based on \selavy\ rms maps), forced fits for non-detections, or forced fits for all epochs.

\subsubsection{Pipeline interface}
Users are able to explore the results from the VAST Pipeline in an interactive manner using the Pipeline class, which contains built-in methods to perform tasks such as transient analyses, plotting lightcurves and querying external resources. This is achieved by using the parquet files that are provided as output by the pipeline, and using data exploration libraries such as \software{Pandas} and \software{vaex}\footnote{\url{https://vaex.io}.}.

\section{UNTARGETED VARIABILITY SEARCH}\label{s_search}
Some of the early results from VAST focus on the time-domain behaviour of samples of known objects selected in radio or other wavebands, for example the search for radio afterglows from \textit{Swift} GRBs that resulted in a detection of the late-time afterglow from GRB~171205A \citep{leung21}. 

In contrast, in this section we present initial results from an untargeted search for highly variable sources and transients in regions 3 and 4 of the VAST Phase I Pilot Survey. This demonstrates the functionality of the VAST pipeline, and shows some of the diversity of variable sources that the full VAST survey will detect.
Analysis of the other survey regions will be presented in subsequent papers.

\subsection{Data analysis}
\label{s_analysis}
We ran the VAST pipeline on the fields from regions 3 and 4, following the process outlined in Section~\ref{s_pipeline}. We then selected compact, isolated sources by excluding sources that satisfied any of the following criteria:
\begin{itemize}
  \item fewer than two measurements (forced or \selavy);
  \item neighbouring sources within 30\,arcseconds;
  \item extended (a ratio of average integrated flux to average peak flux greater than 1.5);
  \item a SNR of $<7$;
  \item negative modulation index (\red{caused by a negative mean flux density}).
\end{itemize}
After applying these criteria, our remaining sample of compact, isolated sources consisted of 155\,071 unique sources each detected in at least one epoch.   This is a very conservative selection.  For instance, requiring no nearby neighbours helps to eliminate imaging artifacts but will also exclude sources in a galaxy with a detectable nucleus or sources in crowded regions.  Therefore the results below should be taken as a lower limit, and future searches should identify more objects even in the same regions of the sky.

\begin{figure}
	\includegraphics[width=\columnwidth]{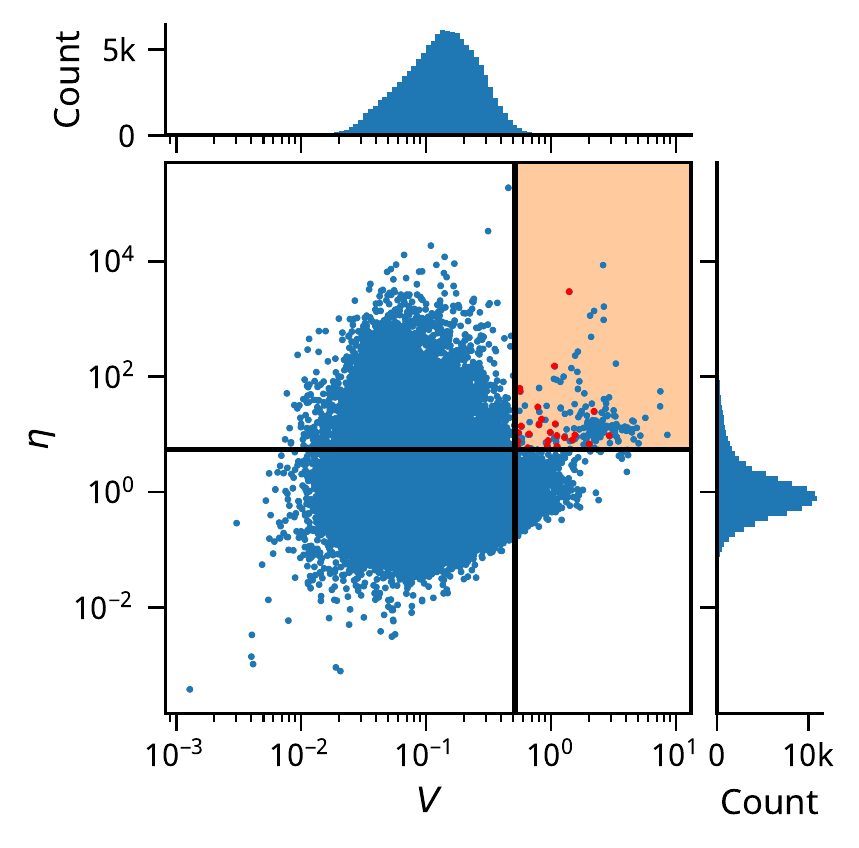}
    \caption{A plot of the two key variability metrics, $V$ and $\eta$. Sources that appear in the shaded top-right quadrant are variable candidates as they exceed the $2\sigma$ thresholds on $V$ and $\eta$ calculated by fitting a Gaussian function to the sigma-clipped distributions of each metric. The thresholds are $V > 0.51$ and $\eta > 5.53$. Sources that have been classified as variables after manual inspection are marked in red.} \label{f_variables}
\end{figure}

Figure~\ref{f_variables} shows the distribution of sources in the two key variability metrics $V$ and $\eta$ (defined in Equations~\ref{e_var2} and \ref{e_var1}). Using $2\sigma$ cutoffs in both of these parameters ($V>0.51$, $\eta > 5.53$) we identified 171 sources as highly variable. These appear in the shaded top right-hand quadrant of Figure~\ref{f_variables}. 

We manually inspected the radio light curves, images and multi-wavelength data for each of these sources, and identified 108 as artefacts near bright sources. Another 14 were in regions of poor data quality, and 21 were detections of marginal significance. This left 28 reliable astronomical transients and variables, which are marked as red in Figure~\ref{f_variables}. We discuss these variable sources in the next section. 

\subsection{Results}\label{s_results}
Our 28 highly variable and transient sources are listed in Table~\ref{t_results} with light curves in Figure~\ref{f_lightcurves} and image cutouts in Figure~\ref{f_cutouts}.
Seven of these sources are known pulsars and seven are radio stars. We discuss each of these classes below. The remaining sources are discussed in Section~\ref{s_unk}.

\subsubsection{Pulsars}
Seven of our highly variable sources are known pulsars: PSRs~J0255$-$5304, J0418$-$4154, J0600$-$5756, J2039$-$5617, J2144$-$5237, J2155$-$5641, and J2236$-$5527.  Pulsars can be variable for intrinsic reasons (nulling, intermittency, e.g., \citealt{backer70,kramer06,wang20}), external reasons (interstellar scintillation; e.g., \citealt{rickett70,stinebring90,bell16,kumamoto21}), or because of absorption by gas in binary systems  \citep[e.g.,][]{broderick16,polzin20,kudale20}.  In fact this variability can help separate pulsars from the background of continuum sources \citep[e.g.,][]{dai17}.

Three of the sources are in binary systems: PSRs J2039$-$5617, J2144$-$5237, and J2236$-$5527.  Of these, PSR~J2039$-$5617 does have eclipses that cause radio pulsations to disappear for half the orbit \citep{corongiu21}.  That source also has significant variability in the pulsed amplitude, which \citet{corongiu21} largely ascribe to scintillation but also argue could be caused by more `exotic' mechanisms (such as absorption by intra-binary gas).  PSR~J2144$-$5237 is a binary with a low-mass companion \citep{bhattacharyya19} that, based on its mass alone, could be a `redback' system. Such systems are often eclipsing \citep{roberts11}, but there is no sign of eclipses in the timing data and so a helium-core white dwarf seems more likely.  PSR~J2236$-$5527 is similarly presumed to have a helium-core white dwarf companion \citep{burgay13}.   Therefore binary companions do not seem to be causing the variability in these systems, although PSR~J2039$-$5617 may merit further analysis.

\begin{figure}
    \centering
    \includegraphics[width=\columnwidth]{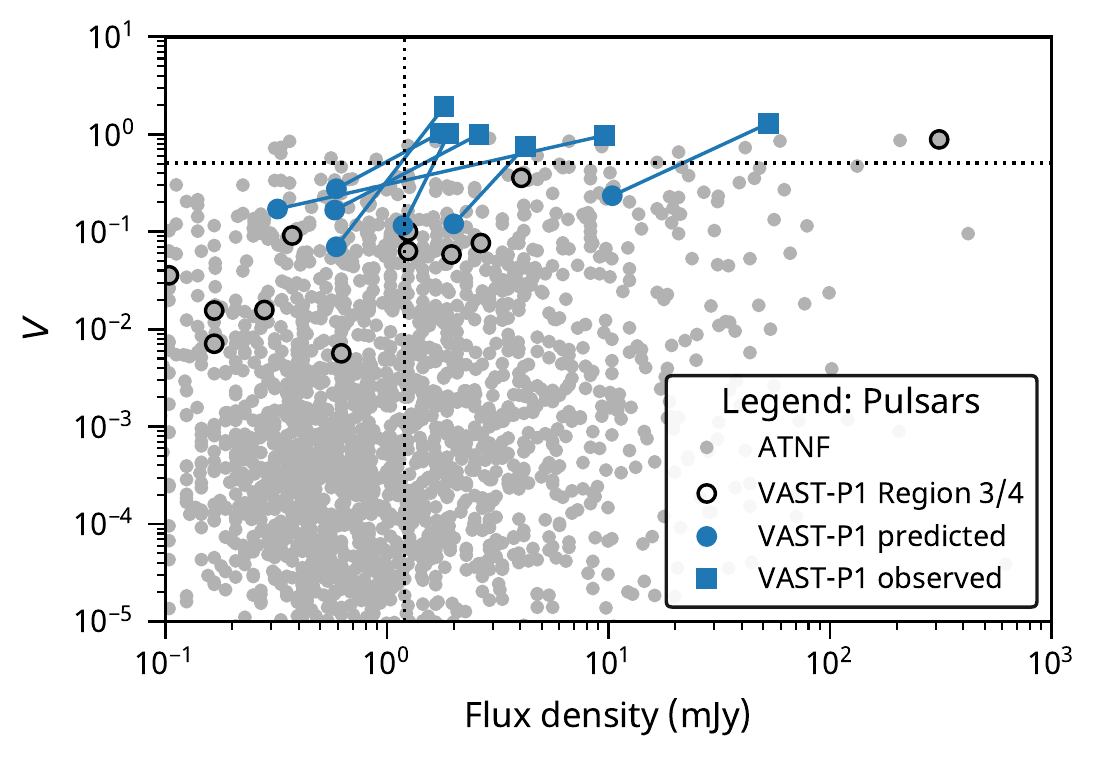}
    \caption{Predicted modulation index (for interstellar scintillation) $V$ versus 888-MHz flux density for pulsars.  We show all of the pulsars in the ATNF pulsar catalogue \citep{manchester16} as grey points, with the flux densities computed from the catalogued values at 1.4\,GHz or 430\,MHz assuming a spectral index of $-1.6$ \citep{jankowski18}.  Predicted modulation index is calculated using the Galactic electron-density model of \citet{yao17} and diffractive scintillation following \citet{lorimer12} and \citet{cordes91}.   The pulsars contained but not detected in VAST-P1 regions 3 and 4 are black outlined circles. The highly variable pulsars identified here are blue circles (based on catalogue values, with the addition of \citealt{corongiu21} for PSR~J2039$-$5617 and \citealt{bhattacharyya19} for PSR~J2144$-$5237) that are connected to their observed properties in VAST-P1 (blue squares).  Rather than mean flux density  we show the maximum flux density, as that governs detectability.  Our thresholds of flux density {$\geq 1.2\,$mJy} (based on the median image rms and a $5\sigma$ detection threshold) and modulation index $\geq0.51$ are the dashed lines.  The black open circle at the right near $300\,$mJy is PSR~J0437$-$4715: it was not identified in our sample because it is below our $V$ threshold (it has $V=0.4$) and it has an unrelated source within the 30\,arcsec neighbour limit.
    }
    \label{fig:pulsars}
\end{figure}

For the other four pulsars, there are no exotic phenomena noted in the literature \citep[e.g.,][]{newton81,xie19,johnston21}.  Therefore, in the absence of any other reason we would expect these detections to be due to variability via refractive interstellar scintillation on longer timescales \citep[e.g.,][]{kumamoto21}, plus diffractive interstellar scintillation on shorter timescales \citep{rickett90,cordes+98} although since not all pulsars have been observed extensively we cannot rule out other effects.  To examine this in more detail, in Figure~\ref{fig:pulsars} we plot for all known pulsars the predicted modulation index for interstellar scintillation versus the flux density, and highlight those identified here. The predictions are far from perfect. We have assumed a single spectral index for our flux density estimates, and a rough electron density model (there is more than an order of magnitude variation in scattering timescale as a function of dispersion measure; \citealt{kumamoto21}) along with a single source velocity for our scintillation estimates. However, the sources identified through variability are among those expected to be brighter and more heavily modulated.  We also highlight the difference between the predicted properties (which will reflect the mean flux densities) and the observed properties, which will be biased to high flux densities and modulation levels because of the nature of these detections.   

The maximum dispersion measure (DM) in regions 3 and 4 is about $30\,{\rm pc\,cm}^{-3}$ \citep{yao17}, which is also consistent with the DMs of the pulsars we identified (they span 14--30$\,{\rm pc\,cm}^{-3}$).  As the scintillation timescale and bandwidth are largely dependent on the DM \citep[e.g.,][]{kumamoto21}, the DM range can be used as a proxy to help identify which other sources will be detected as variable in the remainder of our survey.  

% for future work?
%\todo{how many detected epochs per source?  is this consistent with scintillation?  how many new sources if we increase the number of epochs but keep the fields the same?  vs.\ move to new fields?}

\subsubsection{Stars}\label{s_stars}
Seven of our highly variable sources are identified with known stars.
Two of them (CD--44~1173 and UPM J0409--4435) were found in the circular polarisation search conducted by \citet{pritchard21}, and are discussed there. Our new detections (AB Dor A/C, SCR~J0533--4257, UCAC3~89--412162, LEHPM 2-783 and 2MASS J22414436$-$6119311) are discussed below and the images are shown in Figure~\ref{f_cutouts}.

% the point at which figures get mentioned could be reviewed. figures dont get referenced in order.

\begin{figure*}
    \centering
	\includegraphics[width=\textwidth]{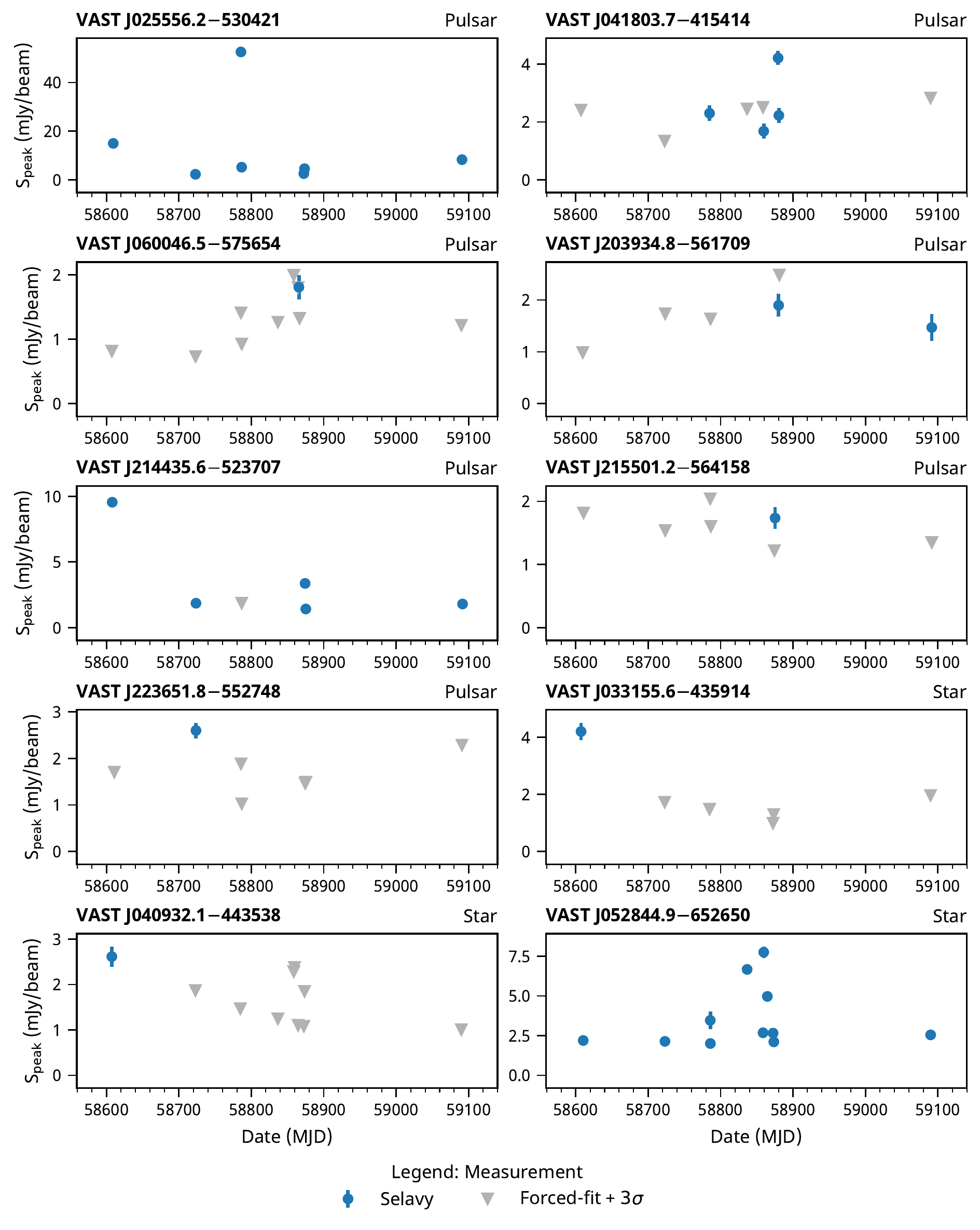}
    \caption{Lightcurves of the variables identified in Regions 3 and 4 of VAST-P1. Blue round points are peak flux density measurements from \selavy{}. Grey triangles are the $3\sigma$ upper-limits of the forced-fitted flux density for images where there was no \selavy{} detection.} 
    \label{f_lightcurves}
\end{figure*}
\begin{figure*}\ContinuedFloat
    \centering
	\includegraphics[width=\textwidth]{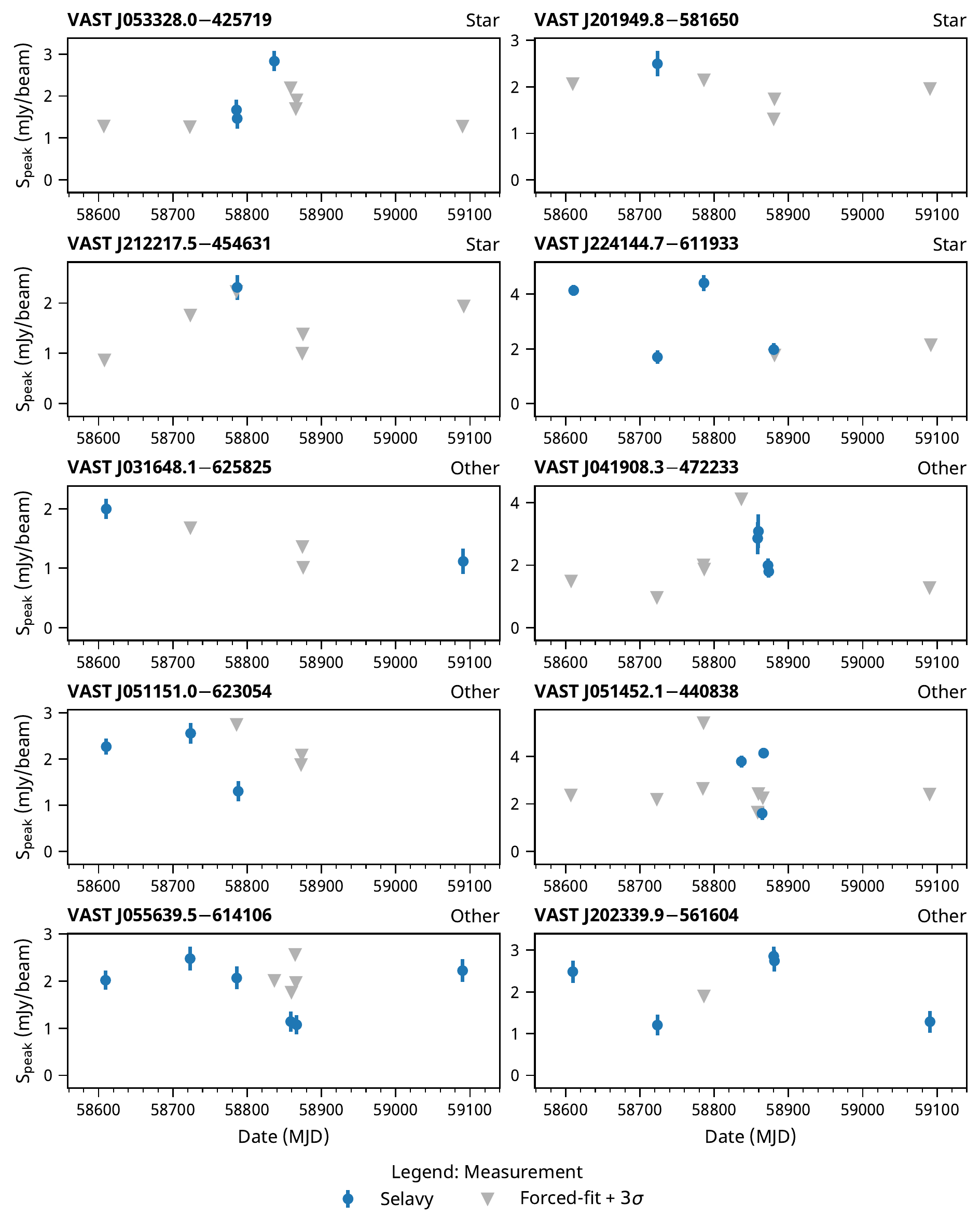}
    \caption{(continued) Lightcurves of the variables identified in Regions 3 and 4 of VAST-P1. Blue round points are peak flux density measurements from \selavy{}. Grey triangles are the $3\sigma$ upper-limits of the forced-fitted flux density for images where there was no \selavy{} detection.} 
\end{figure*}
\begin{figure*}\ContinuedFloat
    \centering
	\includegraphics[width=\textwidth]{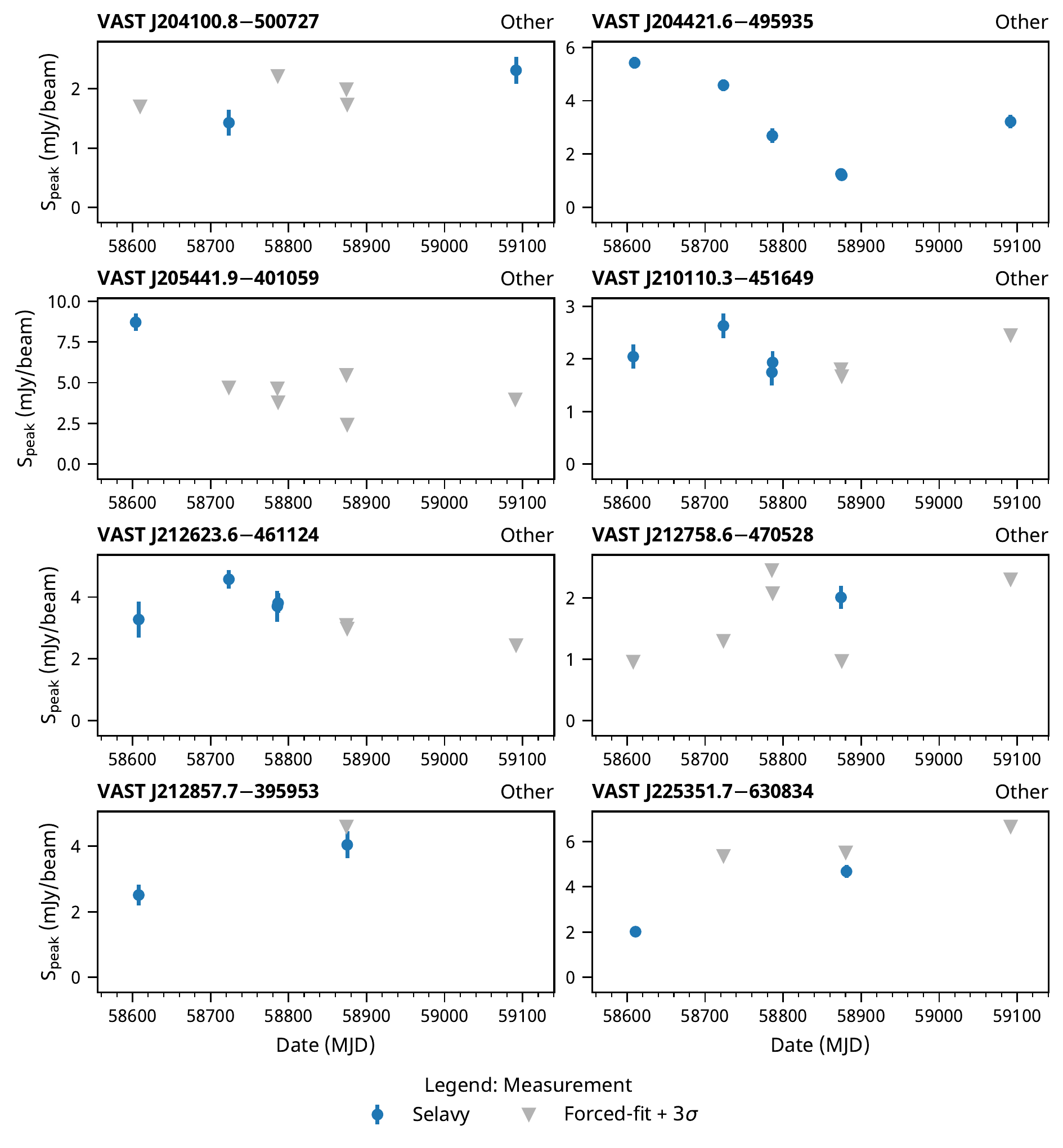}
    \caption{(continued) Lightcurves of the variables identified in Regions 3 and 4 of VAST-P1. Blue round points are peak flux density measurements from \selavy{}. Grey triangles are the $3\sigma$ upper-limits of the forced-fitted flux density for images where there was no \selavy{} detection.} 
\end{figure*}

\begin{figure*}
    \centering
    \includegraphics[width=\textwidth]{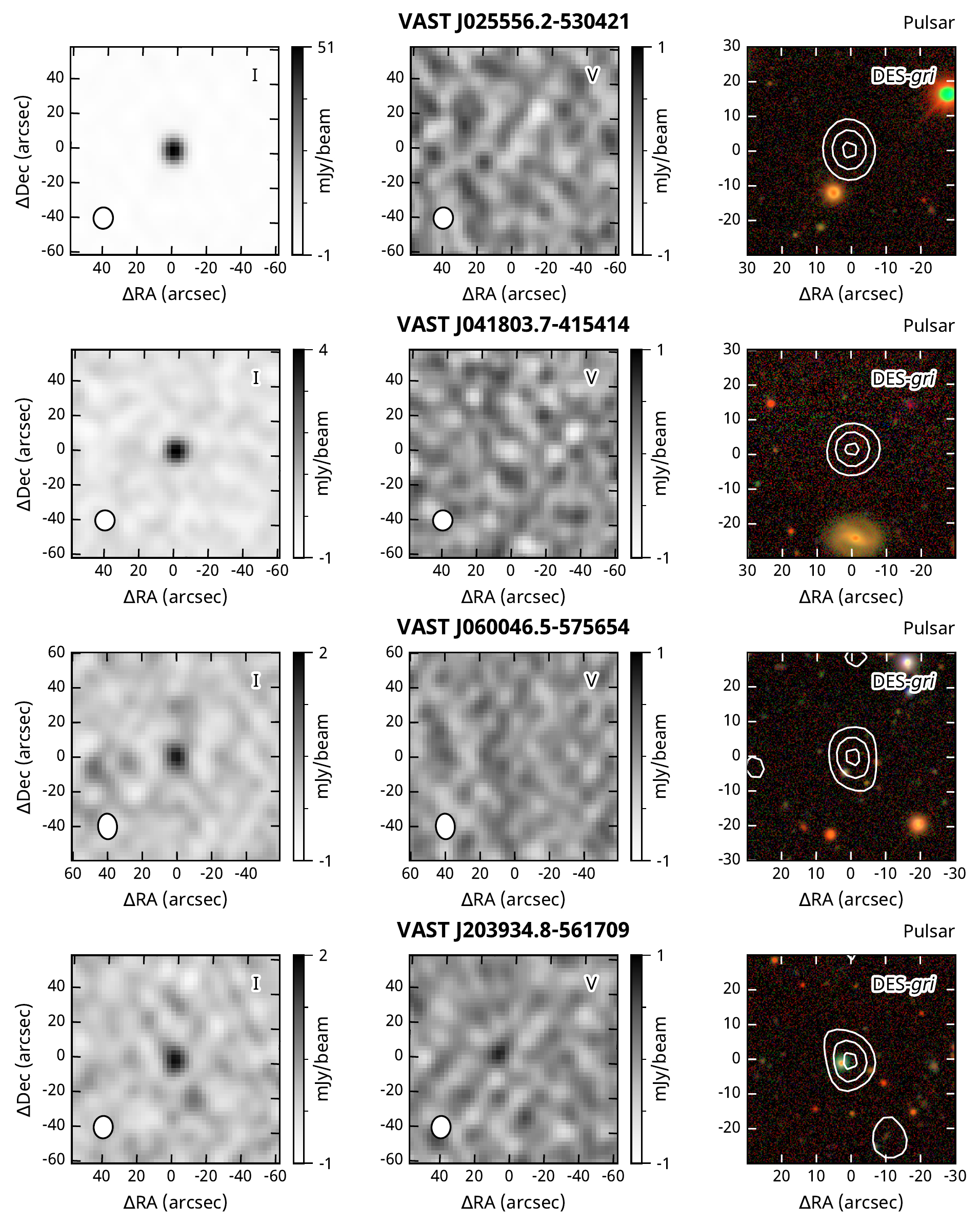}
    \caption{Images of the variable sources identified in this paper. The left panel shows the VAST Stokes I image for the epoch with the maximum flux density. The middle panel shows the Stokes V image for the same epoch where positive flux density corresponds to right handed circular polarisation and negative to left handed. The ellipse in the lower left corner of each radio image shows the FWHM of the restoring beam. The right panels show Stokes I contours at 30, 60, and 90 per cent of the peak Stokes I flux density overlaid on an RGB image of optical data from either DES~DR1 (DES) or SkyMapper \citep[SM;][]{onken19} with red=\textit{i}-band, green=\textit{r}-band, blue=\textit{g}-band. All images have been centred on a frame aligned with the position of the radio source. Optical data have been astrometrically corrected to the VAST epoch according to the proper motion of the target when available.}
    \label{f_cutouts}
\end{figure*}
\begin{figure*}\ContinuedFloat
    \centering
    \includegraphics[width=\textwidth]{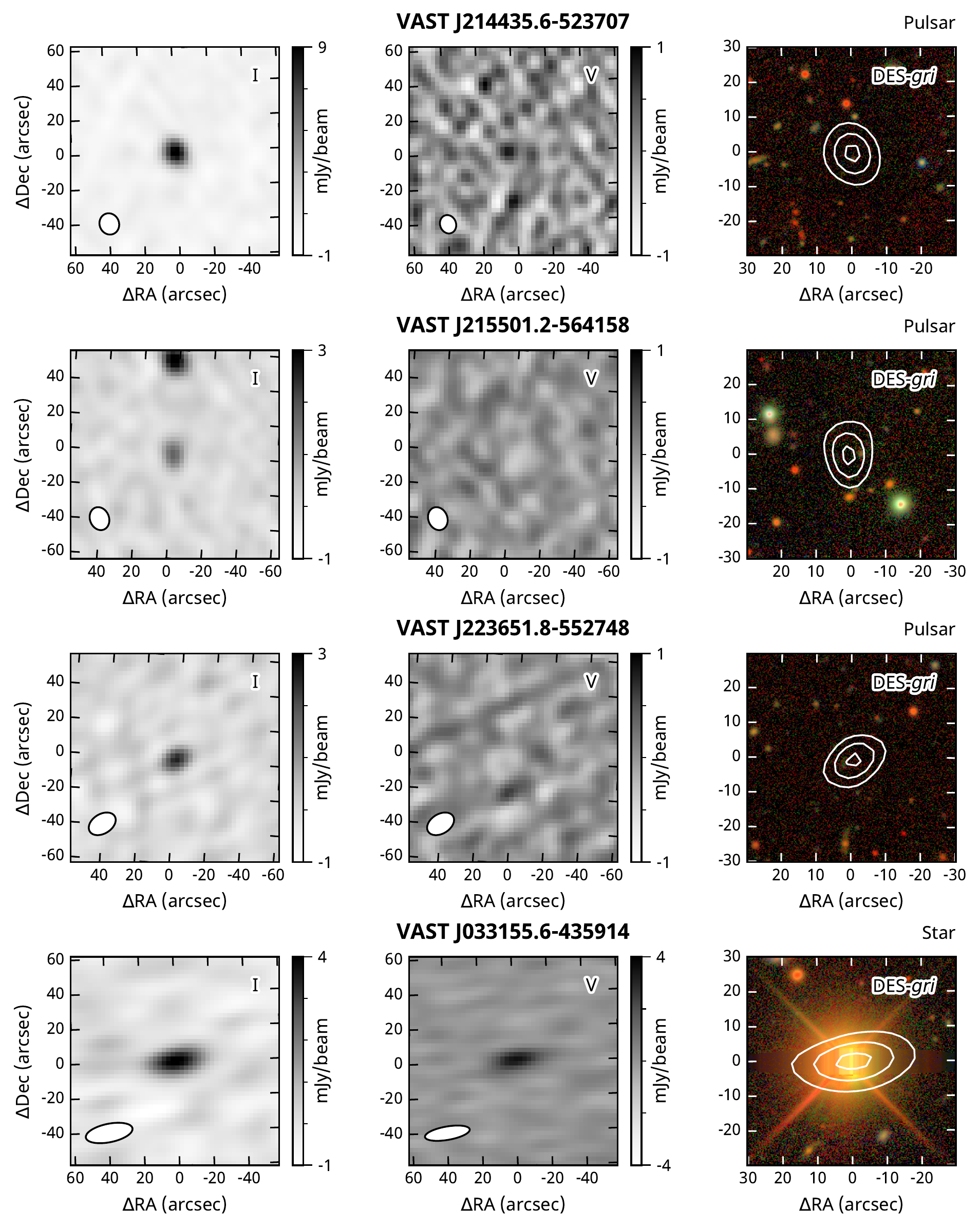}
    \caption{(continued) Images of the variable sources identified in this paper. The left panel shows the VAST Stokes I image for the epoch with the maximum flux density. The middle panel shows the Stokes V image for the same epoch where positive flux density corresponds to right handed circular polarisation and negative to left handed. The ellipse in the lower left corner of each radio image shows the FWHM of the restoring beam. The right panels show Stokes I contours at 30, 60, and 90 per cent of the peak Stokes I flux density overlaid on an RGB image of optical data from either DES~DR1 (DES) or SkyMapper \citep[SM;][]{onken19} with red=\textit{i}-band, green=\textit{r}-band, blue=\textit{g}-band. All images have been centred on a frame aligned with the position of the radio source. Optical data have been astrometrically corrected to the VAST epoch according to the proper motion of the target when available.}
\end{figure*}
\begin{figure*}\ContinuedFloat
    \centering
    \includegraphics[width=\textwidth]{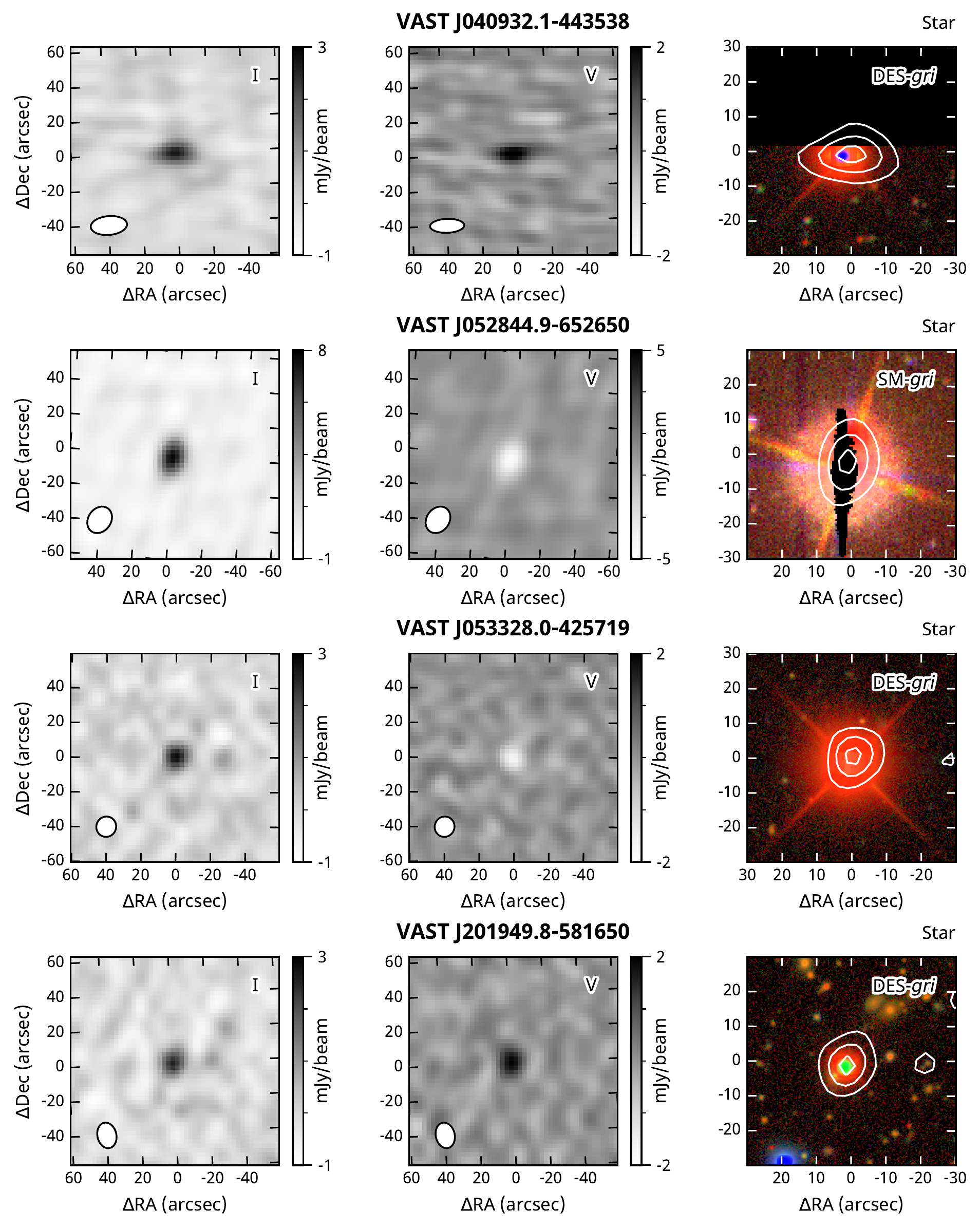}
    \caption{(continued) Images of the variable sources identified in this paper. The left panel shows the VAST Stokes I image for the epoch with the maximum flux density. The middle panel shows the Stokes V image for the same epoch where positive flux density corresponds to right handed circular polarisation and negative to left handed. The ellipse in the lower left corner of each radio image shows the FWHM of the restoring beam. The right panels show Stokes I contours at 30, 60, and 90 per cent of the peak Stokes I flux density overlaid on an RGB image of optical data from either DES~DR1 (DES) or SkyMapper \citep[SM;][]{onken19} with red=\textit{i}-band, green=\textit{r}-band, blue=\textit{g}-band. All images have been centred on a frame aligned with the position of the radio source. Optical data have been astrometrically corrected to the VAST epoch according to the proper motion of the target when available.}
\end{figure*}
\begin{figure*}\ContinuedFloat
    \centering
    \includegraphics[width=\textwidth]{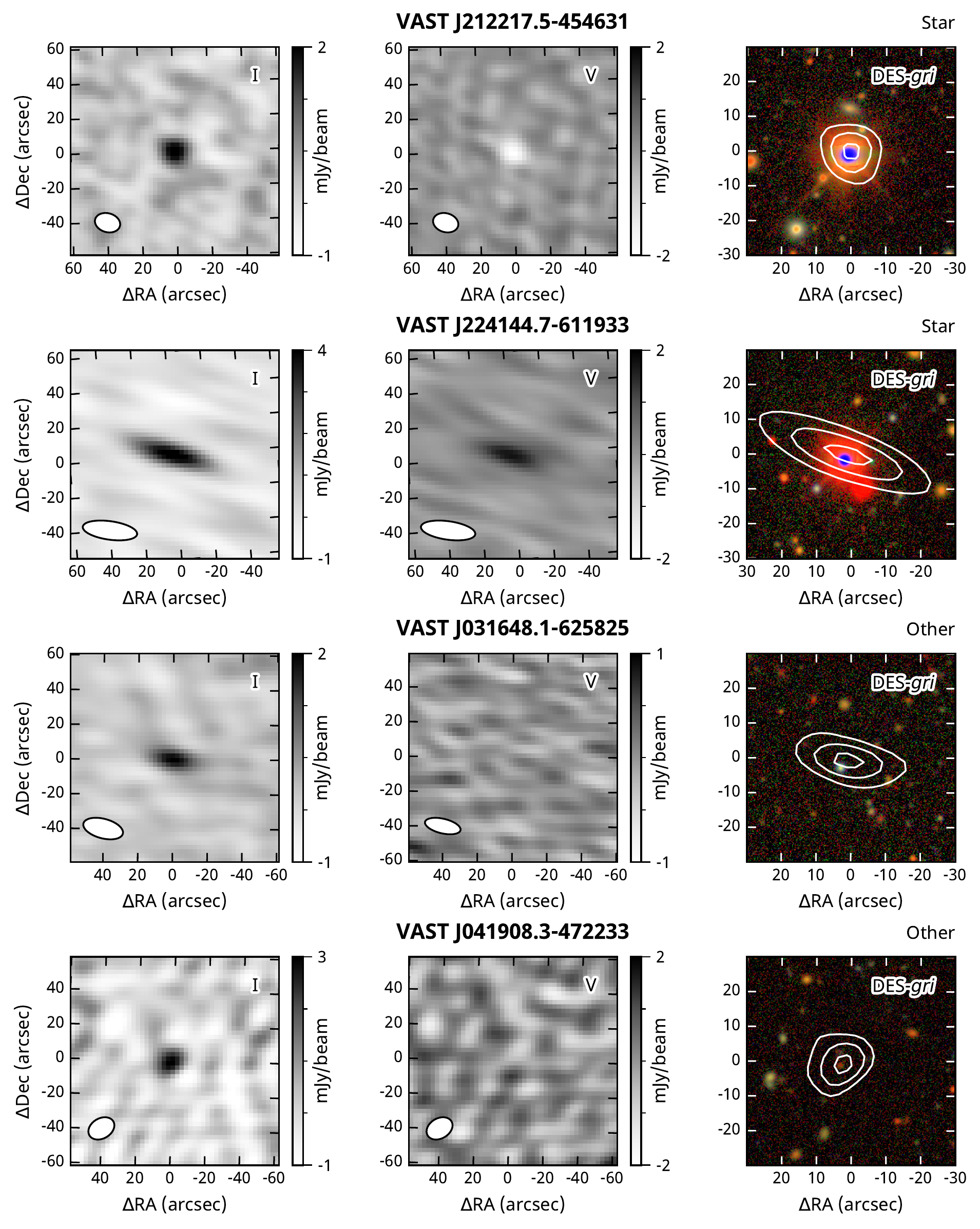}
    \caption{(continued) Images of the variable sources identified in this paper. The left panel shows the VAST Stokes I image for the epoch with the maximum flux density. The middle panel shows the Stokes V image for the same epoch where positive flux density corresponds to right handed circular polarisation and negative to left handed. The ellipse in the lower left corner of each radio image shows the FWHM of the restoring beam. The right panels show Stokes I contours at 30, 60, and 90 per cent of the peak Stokes I flux density overlaid on an RGB image of optical data from either DES~DR1 (DES) or SkyMapper \citep[SM;][]{onken19} with red=\textit{i}-band, green=\textit{r}-band, blue=\textit{g}-band. All images have been centred on a frame aligned with the position of the radio source. Optical data have been astrometrically corrected to the VAST epoch according to the proper motion of the target when available.}
\end{figure*}
\begin{figure*}\ContinuedFloat
    \centering
    \includegraphics[width=\textwidth]{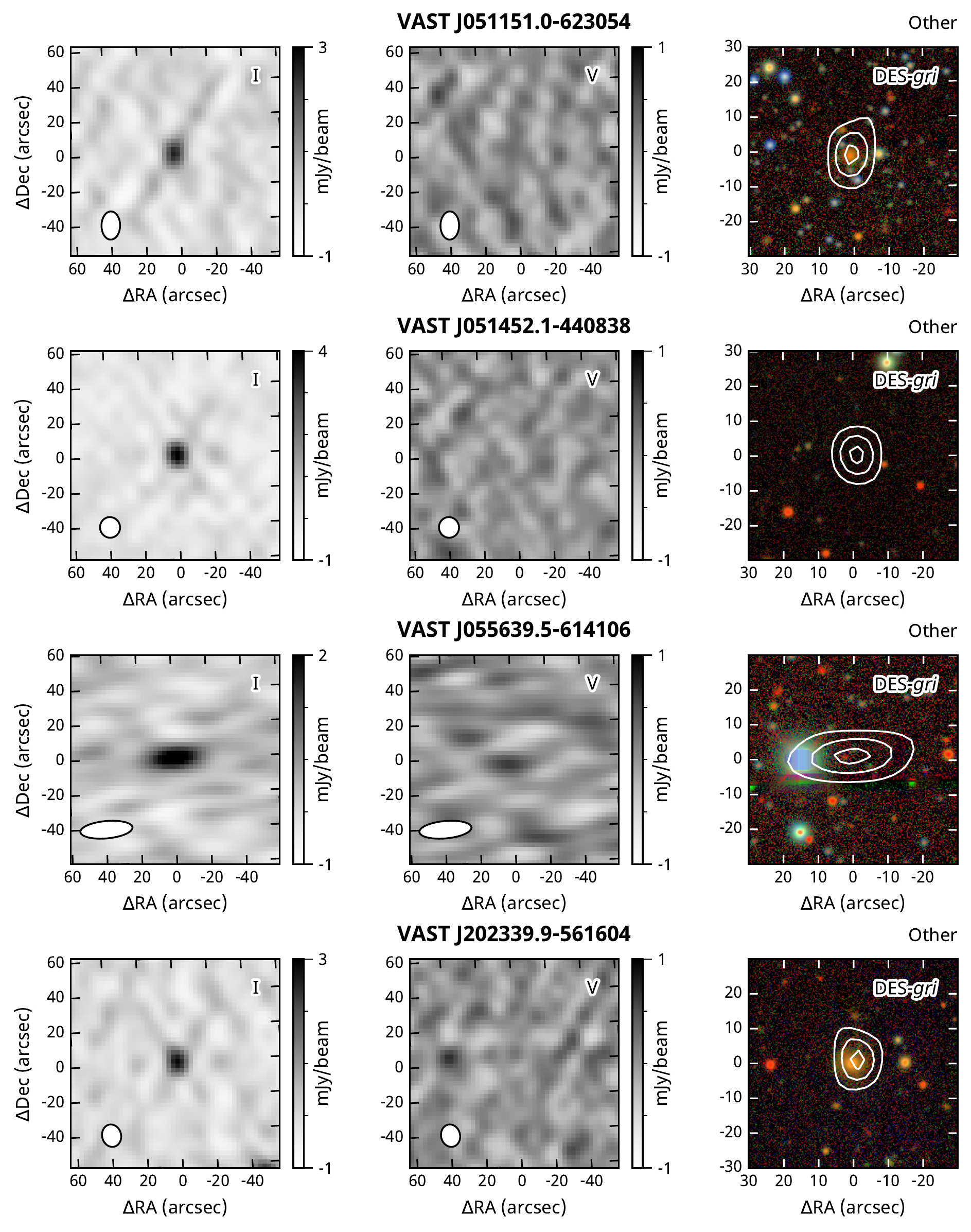}
    \caption{(continued) Images of the variable sources identified in this paper. The left panel shows the VAST Stokes I image for the epoch with the maximum flux density. The middle panel shows the Stokes V image for the same epoch where positive flux density corresponds to right handed circular polarisation and negative to left handed. The ellipse in the lower left corner of each radio image shows the FWHM of the restoring beam. The right panels show Stokes I contours at 30, 60, and 90 per cent of the peak Stokes I flux density overlaid on an RGB image of optical data from either DES~DR1 (DES) or SkyMapper \citep[SM;][]{onken19} with red=\textit{i}-band, green=\textit{r}-band, blue=\textit{g}-band. All images have been centred on a frame aligned with the position of the radio source. Optical data have been astrometrically corrected to the VAST epoch according to the proper motion of the target when available.}
\end{figure*}
\begin{figure*}\ContinuedFloat
    \centering
    \includegraphics[width=\textwidth]{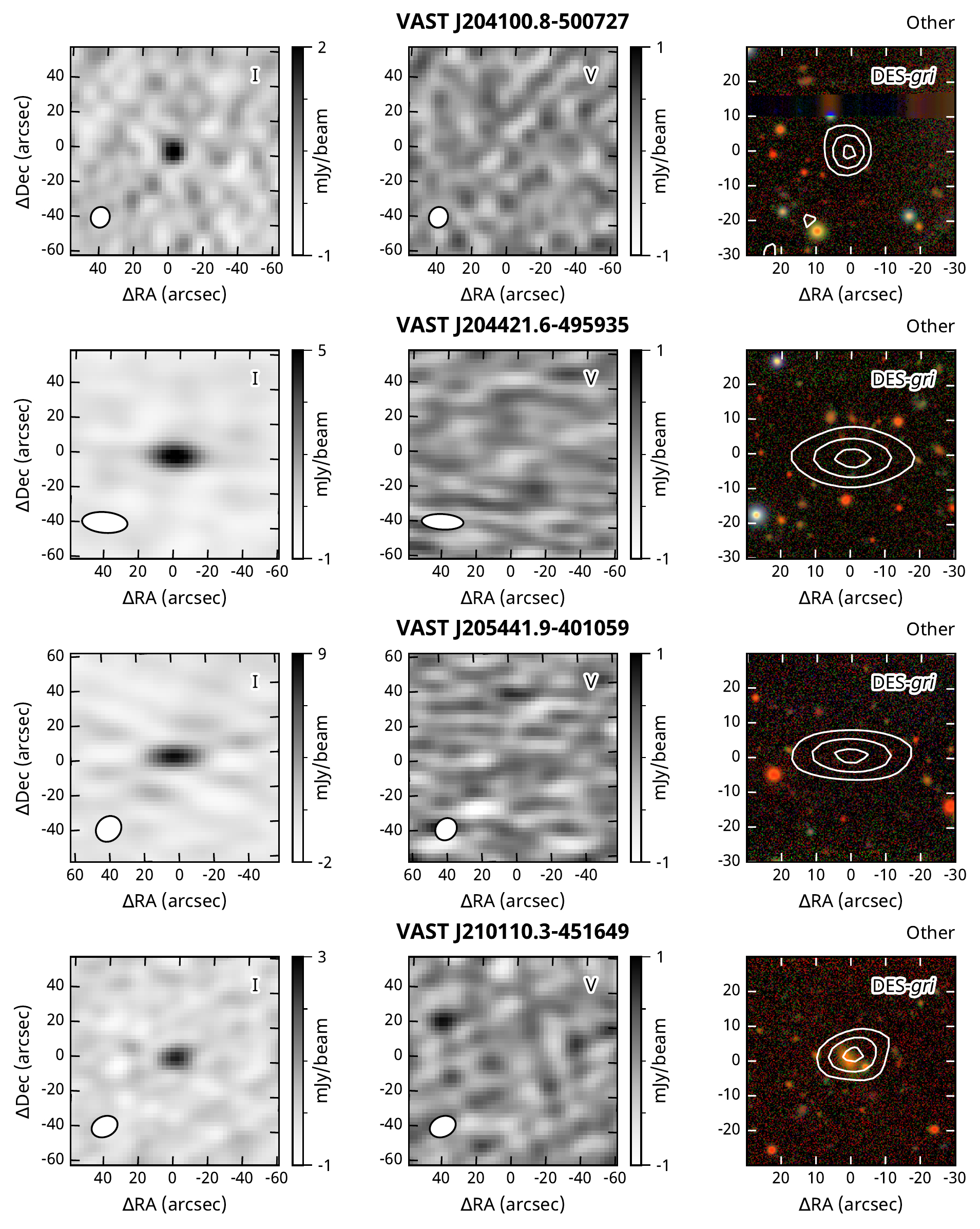}
    \caption{(continued) Images of the variable sources identified in this paper. The left panel shows the VAST Stokes I image for the epoch with the maximum flux density. The middle panel shows the Stokes V image for the same epoch where positive flux density corresponds to right handed circular polarisation and negative to left handed. The ellipse in the lower left corner of each radio image shows the FWHM of the restoring beam. The right panels show Stokes I contours at 30, 60, and 90 per cent of the peak Stokes I flux density overlaid on an RGB image of optical data from either DES~DR1 (DES) or SkyMapper \citep[SM;][]{onken19} with red=\textit{i}-band, green=\textit{r}-band, blue=\textit{g}-band. All images have been centred on a frame aligned with the position of the radio source. Optical data have been astrometrically corrected to the VAST epoch according to the proper motion of the target when available.}
\end{figure*}
\begin{figure*}\ContinuedFloat
    \centering
    \includegraphics[width=\textwidth]{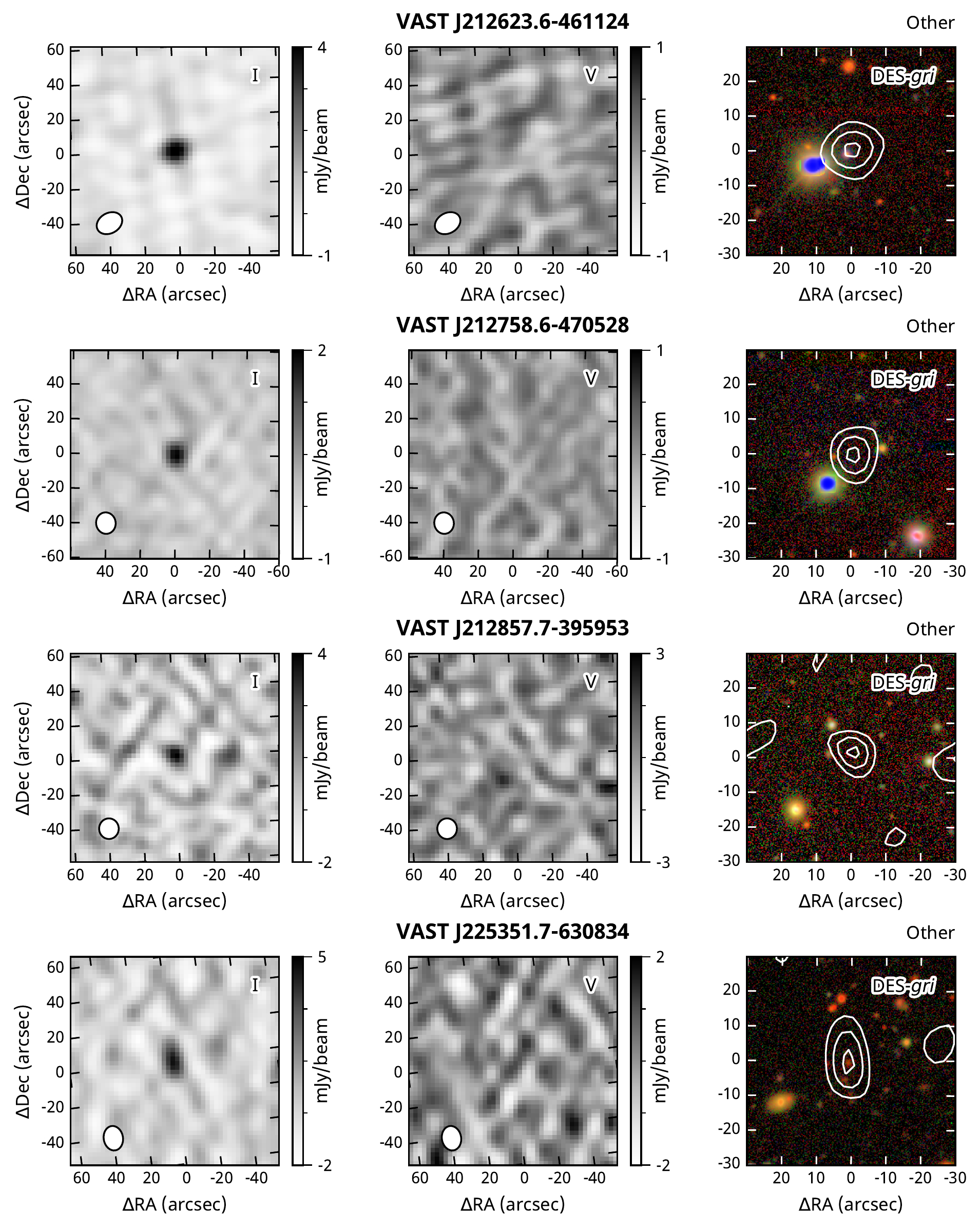}
    \caption{(continued) Images of the variable sources identified in this paper. The left panel shows the VAST Stokes I image for the epoch with the maximum flux density. The middle panel shows the Stokes V image for the same epoch where positive flux density corresponds to right handed circular polarisation and negative to left handed. The ellipse in the lower left corner of each radio image shows the FWHM of the restoring beam. The right panels show Stokes I contours at 30, 60, and 90 per cent of the peak Stokes I flux density overlaid on an RGB image of optical data from either DES~DR1 (DES) or SkyMapper \citep[SM;][]{onken19} with red=\textit{i}-band, green=\textit{r}-band, blue=\textit{g}-band. All images have been centred on a frame aligned with the position of the radio source. Optical data have been astrometrically corrected to the VAST epoch according to the proper motion of the target when available.}
\end{figure*}

\src{VAST J052644.9$-$652650} is identified as AB~Dor, a quadruple system composed of a K1V pre-main sequence star + M5.5V and M3.5 + M4.5 pair of binary systems \citep{Close2007}. The pre-main sequence star AB~Dor A, is by far the brightest star in the system; it has a rotation period of 0.514 d, and shows optical super-flares 
\citep{Schmitt2019} and long-term X-ray variability \citep{Lalitha2013}. VLBI radio observations at
\SI{1.4}{\giga\hertz}, \SI{8.4}{\giga\hertz}, and \SI{22.3}{\giga\hertz} have revealed milliarcsecond 
radio structure \citep{Climent2020}. The source is detected in VAST-P1 in 11 epochs in Stokes I with a 
maximum flux density of 7.8\,mJy\,beam$^{-1}$, and in three epochs in Stokes V
showing both right- and left-handed circular polarisation with circular polarisation fraction ranging 
from $52\%$ to $60\%$. We speculate that the three events with high circular polarisation are associated with known prominence formation in young active stars such as AB Dor, HK Aqr, BO Mic and PZ Tel \citep{Doyle1990,Barnes2001,Jardine2001}. AB Dor lies in the Transiting Exoplanet Survey Satellite (TESS; \citealt{Ricker2015}) southern continuous viewing zone, allowing near continuous optical photometry at 2 min cadence extending over a year \citep{IoannidisSchmitt2020}. 

\src{VAST J053328.0$-$425719} is identified as SCR~J0533--4257, an M4.5 dwarf \citep{Riaz2006}. Ultraviolet activity \citep{Schneider2018} and X-ray flares \citep{Fuhrmeister2003} have been observed from this star, though there are no previous radio detections reported in the literature. The source is detected in three epochs. The maximum flux density is $2.83\pm0.24$\,mJy\,beam$^{-1}$. The circular polarisation fraction ranges between $33\%$ and $57\%$. It was observed using TESS (TIC 302962949) in 2\,min cadence during
sectors 5 and 6 and in 20\,sec cadence in sector 33 (total duration
67.9\,d). The TESS data show evidence for modulations at
periods of 1.25 and 0.46\,d, hinting it may be a binary. Optical flares are
frequently seen, most of which have a peak amplitude of a few $\sim$0.01\,mag but up to 0.2\,mag on one occasion.

\src{VAST J201949.8$-$581650} is identified as LEHPM~2--783, an M6 dwarf \citep{Riaz2006} with high proper motion ($\mu_{\alpha}$=--23.4, $\mu_{\delta}$=--341.1 mas yr$^{-1}$) \citep{Gaia2018}. There are no previous radio detections reported in the literature. The source is only detected in a single epoch with a flux density of $2.50\pm0.27$\,mJy\,beam$^{-1}$ and a circular polarisation fraction of $77\%$. 
It was observed using TESS (TIC 387220832) in 2\,min cadence during
sectors 13 and 27 (total duration 50.4\,d). The TESS sector 13 data
shows evidence for a period at 6.7\,d whilst the sector 27 data suggest
a longer period of 11.6\,d. It is flare-active, with several optical flares
having amplitudes greater than 0.5\,mag. It has a location on the \gaia\
colour-magnitude diagram red-ward of the main sequence, indicating it maybe a binary or a
young star. The high circulation polarisation of 77\% could point towards auroral magnetospheric activity, similar to that seen by \citet{zic19} in UV Ceti, a star of similar spectral type.

\src{VAST J212217.5$-$454631} is identified as UCAC3 89--412162, which has a distance of 31.6 pc and with a \gaia\ colour $(\mathrm{BP}-\mathrm{RP}) =2.49$ and absolute magnitude MG $=10.0$ \citep{Gaia2018} is consistent with an M3V spectral type \citep{pecaut13}. There are no previous radio detections reported in the literature. This source is also detected in a single epoch with a flux density of $2.31\pm0.25$\,mJy\,beam$^{-1}$ and a circular polarisation fraction of $87\%$.
It was observed using TESS (TIC 200645542) in 2\,min cadence during
sectors 27 and 28 (total duration 41.1\,d). There is no evidence for a
periodic modulation in the optical light curve or flaring activity. The lack of flaring activity in this object is consistent with reduced or minimum spot coverage. This of course does not mean a lack of flaring, e.g., \citet{Schmitt2019} suggested that large flares only have $\sim$14\% of their energy in the TESS passband. The fact that we detect a radio outburst could suggest that this object may have flares that have a peak flux in the blue, since the contrast over the background emission is much higher at shorter wavelengths and the TESS wavelength band in centred at 7865 \AA.

\src{VAST J224144.7$-$611933} is associated with 2MASS~J22414436$-$6119311. It has a high proper motion ($\mu_{\alpha}=150.2$, $\mu_{\delta}=-87.9$ mas yr$^{-1}$, from \gaia), a colour ($\mathrm{BP}-\mathrm{RP}=2.81$) consistent with a M3.5V spectral type and has a distance of 28.42$\pm$0.06 pc \citep{gaia20}. It was observed in TESS sectors 1, 27 and 28 (TIC 232064182) in 2\,min cadence and shows evidence for a periodic modulation on a period of 0.721\,d. Short duration optical flares are seen with amplitudes typically of a few percent, but on one occasion with an amplitude of $\sim$60 percent.

All of the stars in our sample show radio emission with high fractional circular polarisation ($>30\%$), making this a useful tool for distinguishing these sources from extragalactic transients. They have optical magnitudes between V=17.17 and V=10.9, and span a wide range in brightness and colour. With the exception of SCR~J0533$-$4257, these stars have high proper motions ($>100$ mas/yr) suggesting that in the absence of an optical identification, VLBI proper motion measurements are another way in which they can be distinguished from extragalactic sources.

\subsubsection{Other variables}\label{s_unk}
We found 14 highly variable sources that were not associated with a known pulsar or star. 
For these we searched for optical or infrared counterparts in several \wise\ catalogs, {\it \gaia} EDR3 \citep{gaia20} and the Dark Energy Survey DR1. Proper motions from \gaia\ were applied if available. Photometric redshifts were retrieved from the Photometric Redshifts for the Legacy Surveys catalogue \citep{zhou21} when available. For sources that have unclear \wise\ counterparts due to resolution limits, we used forced-photometry measurements from the unWISE images \citep[for W1 and W2]{lang14,meisner16,meisner17} and \wise\ images (for W3) provided in the DESI Legacy Imaging Surveys DR8 catalogue \citep{dey19}. 

\begin{figure}
    \centering
    \includegraphics[width=\columnwidth]{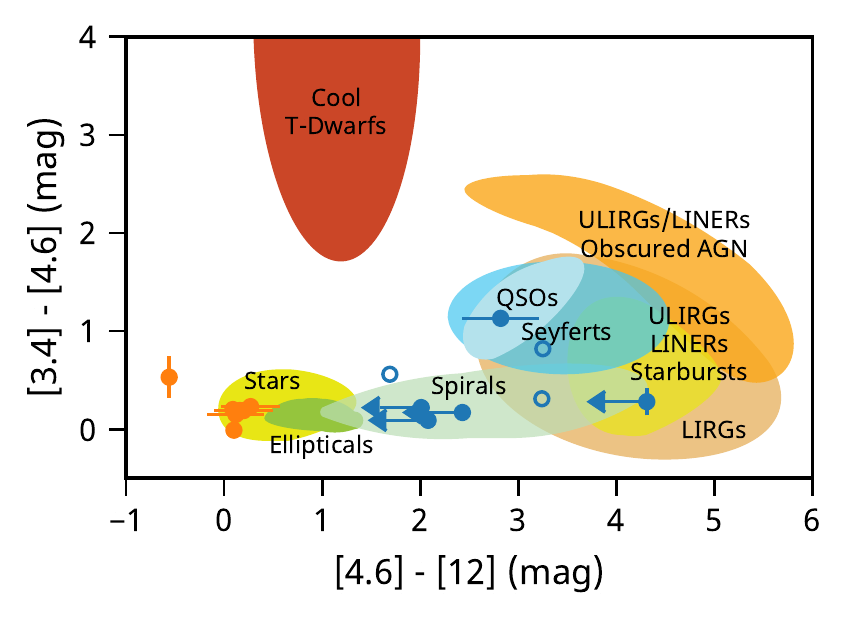}
    \caption{\wise\ colours of the variable sources identified in this paper plotted on top of the classification regions from \citet{wright10}. Sources classified as stars are shown in orange, and those classified as other variables are shown in blue. Sources that did not have a clear counterpart in \wise\ but had unWISE forced-photometry available in the DESI Legacy Imaging Surveys DR8 catalogue are shown with open markers.}
    \label{f_wise}
\end{figure}

None of these 14 sources have a detection in circular polarisation within the VAST-P1 data, although many of the limits are not very constraining, with $3\sigma$ Stokes V/I fractional polarisation limits between $11\%$ and $56\%$. 
The multi-wavelength properties of these objects are summarised in Table~\ref{t_multi}. Sources with a WISE counterpart (including some of the stars discussed in Section~\ref{s_stars}) are plotted on a WISE colour-colour plot in Figure~\ref{f_wise}.

\begin{table*}
\centering
\caption{Multiwavelength properties of the 14 variable sources not identified as known stars or pulsars; see Section~\ref{s_unk}. We provide the \wise\ cross-ID, or for sources with unclear \wise\ counterparts, the unique LS object ID from the DESI Legacy Imaging Surveys DR8 catalogue. The median of the photometric redshift probability distributions are provided when available from \citet{zhou21} along with the limits of the 95\% confidence interval.}
\label{t_multi}
\begin{tabular}{llcc}
\hline
Source Name & Cross-ID & photo-z & Class \\
\hline
VAST J031648.1$-$625825 & Gaia EDR3 4722095986792487296 & 1.16 (0.81, 1.50) & QSO \\
                        & LS-DR8 8796095391534556       & & \\
                        & WISEA J031648.22$-$625825.5   & & \\
VAST J041908.3$-$472233 & LS-DR8 8796098726269814       & 1.06 (0.82, 1.28) & Galaxy \\
                        & WISEA J041908.43$-$472233.1   & & \\
VAST J051151.0$-$623054 & LS-DR8 8796095481781222       & -- & Galaxy \\
                        & WISEA J051151.13$-$623053.6   & & \\
VAST J051452.1$-$440838 & --                            & -- & ? \\
VAST J055639.5$-$614106 & LS-DR8 8796095615927489       & 0.83 (0.39, 1.12) & Galaxy \\
VAST J202339.9$-$561604 & Gaia EDR3 6469160387838360704 & 0.28 (0.27, 0.31) & Galaxy\\
                        & LS-DR8 8796096719948142       & & \\
                        & WISEA J202340.02$-$561604.8   & & \\
VAST J204100.8$-$500727 & --                            & -- & ? \\
VAST J204421.6$-$495935 & --                            & -- & ? \\
VAST J205441.9$-$401059 & --                            & -- & ? \\
VAST J210110.3$-$451649 & LS-DR8 8796099359737883       & 0.48 (0.41, 0.50) & Galaxy \\
                        & WISEA J210110.42$-$451648.9   & & \\
VAST J212623.6$-$461124 & Gaia EDR3 6575524939389655808 & 2.32 (0.19, 2.68) & Seyfert \\
                        & LS-DR8 8796099096154771       & & \\
VAST J212758.6$-$470528 & --                            & -- & ? \\
VAST J212857.7$-$395953 & --                            & -- & ? \\
VAST J225351.7$-$630834 & LS-DR8 8796095383739344       & 1.00 (0.86, 1.16) & Galaxy \\
\hline 
\end{tabular}
\end{table*}

Two of the objects are associated with AGN. VAST~J031648.1--625825 is coincident with the IR source WISEA~J031648.22--625825.5 (and Gaia source EDR3~472209598679248729). The \wise\ colours of this object (see Figure~\ref{f_wise}) suggest it is a QSO with a photometric redshift\footnote{Redshifts are quoted as the median of the photo-$z$ probability distribution with the 95\% confidence interval.} of $z = 1.16\:(0.82, 1.50)$. There are only two detections in VAST-P1.
VAST J212623.6-461124 is coincident with the \gaia\ source EDR3~657552493938965580, and unWISE colours suggest this is a Seyfert galaxy. There are four detections of this source in VAST-P1.
Both of these \gaia\ sources have parallaxes and proper motions consistent with zero. In the absence of further information it is not possible to determine the cause of the observed variability, but it can likely be explained by scintillation or intrinsic AGN variability.

Six of the sources are associated with galaxies, based on their \wise\ or DES DR1 identifications. None of these are offset from the central positions of their optical or infrared counterparts and so could be due to nuclear emission, but the limited angular resolution means we cannot rule out other sorts of transients (supernovae, GRBs). Therefore, we also searched for possible associations with known transients at other wavelengths. We checked against catalogues of GRBs detected by the International Gamma-Ray Astrophysics Laboratory \citep[INTEGRAL;][]{Mereghetti2003}, the \textit{Swift} Burst Alert Telescope \citep[BAT;][]{Lien2016} and the \textit{Fermi} Large Area Telescope \citep[LAT;][]{Ajello2019} and Gamma-ray Burst Monitor \citep[GBM;][]{vonKienlin2020}. Possible associations with known supernovae were checked against the Open Supernova Catalog\footnote{\url{https://sne.space/}.} \citep{guillochon2017} and supernovae reported in the Transient Name Server\footnote{\url{https://www.wis-tns.org/}.}. None of the sources are associated with known GRBs or supernovae. Further monitoring during VAST-P2 may provide more insight into the origin of the variability in these objects.

The remaining six sources have no obvious optical/IR counterparts. None are detected in circular polarisation (although again the limiting ratios are not very constraining, with $3\sigma$ Stokes V/I ratios ranging between 24\% and 56\%). 
Again, none of these sources are associated with known GRBs or supernovae. 
With the exception of VAST~J204421.6$-$495935, none exhibit a light curve with a temporal rise or decay that is physically consistent with an uncatalogued GRB afterglow or radio supernova.
Further observations of these sources are ongoing.

We used the lightcurves and redshifts (where available) to compute brightness temperatures for all of the sources in Table~\ref{t_multi} (as in \citealt{pietka15}, \citealt{stewart16}, \citealt{bhandari18}, \citealt{bell19}, and others).  For a timescale, we took the minimum timescale over which the two-point modulation index \citep[e.g.,][]{mooley16} exceeded 0.5.  Given the sparsity of our temporal sampling, these timescales do not have good resolution, but range from $\sim 1-2$\,d (e.g., VAST~J051452.1$-$440838) to 266\,d (VAST~J212857.7$-$395953). Using the photometric redshifts (or $z=1$ when not available) we found brightness temperatures of $10^{13}\,$K for the most slowly-varying sources to $>10^{18}\,$K for the most quickly-varying.  However, we do not believe those timescales to be intrinsic to the sources, as they would violate the Compton catastrophe limit \citep{Kellermann1969} or the stricter equipartition brightness temperature limit \citep{readhead94}.  Instead we look at three possibilities: extrinsic variability, Doppler boosting, or coherent emission.

For extrinsic variability, we  used NE2001 \citep[][]{2002astro.ph..7156C} to calculate the scintillation transition frequency, $\nu_0$ for extragalactic sources along the lines of sight contained by regions 3 and 4. Both regions have similar values with the values in region 3 spanning $7.8 - 12.3$\,GHz and the values in region 4 spanning $8.1 - 14.1$\,GHz with median values of 9.2\,GHz and 9.8\,GHz respectively. This quantity can then be used to estimate the expected level of variability caused by refractive scintillation using the scaling relations of \citet{1998MNRAS.294..307W}. We find that compact sources in both regions will exhibit $\sim 25\%$ variability at 888\,MHz due to refractive scintillation on timescales of $\sim 10$ days (the all-sky model for refractive scintillation from \citealt[][]{hancock19} predicts a similar timescale, but a mean modulation index about a factor of two higher than that predicted from NE2001).  In most cases this is reasonably consistent with what we observe, although as we are selecting the most variable sources we are likely probing the tail of the modulation-index distribution.  

However, there are sources that seem inconsistent with this interpretation.  For instance, VAST~J051452.1$-$440838 and VAST~J212758.6$-$470528, which vary by a factor of $>2$ on timescales as fast as 1\,d. If such variability were to be intrinsic and not violate the Compton catastrophe limit, then both sources would be nearby ($\lesssim$5 Mpc); one may have then expected that a counterpart should be detectable at other wavelengths. Instead, in these cases, diffractive scintillation (which has higher modulation indices and faster variability; e.g., \citealt{1998MNRAS.294..307W} and \citealt[][]{macquart06}) may be operating if the source size is very compact.  Or, we may be seeing a relativistic jet beamed toward us with a Doppler factor of $\sim 100$ \citep[e.g.,][]{readhead94,1999ApJ...521..493L}. Finally, some of the the sources without optical counterparts may be coherently-emitting Galactic objects such as pulsars, where diffractive scintillation is expected and intrinsic variability can be very fast.  For most of regions 3 and 4 the most sensitive pulsar survey is the high-latitude portion of the High Time Resolution Universe (HTRU) survey \citep{keith10}, which is sensitive to flux densities as low as 0.2\,mJy at 1.4\,GHz for low DMs.  This is a factor of 10 below the flux densities of our sources, but especially if they are scintillating strongly or otherwise varying, previous surveys may have missed them.
Further observations to probe the spectral energy distribution, small-scale structure, and precise position of the varying source will help distinguish between these scenarios, as can optical spectroscopy to search for signatures of AGN (in those classified as `Galaxy' in Table~\ref{t_multi}) and pulsar searches (for those classified as `?').  

\begin{landscape}
\begin{table}
  \caption{Highly variable sources identified in the VAST-P1 regions 3 and 4. The coordinate of each source is given as the weighted average of all Selavy detections, where the weight is the inverse square of the positional error. $\sigma_\text{pos}$ is the averaged positional uncertainty. $\eta$ and $V$ are the variability parameters described in the text. nE gives the number of epochs (observations) that cover the source location. nD gives the number of detections. $|{\rm V}|/{\rm I}$ is the ratio of Stokes V to Stokes I flux density measured in the epoch for which this is a maximum, or the most constraining 3 sigma upper limit in the case of non-detections in Stokes V.}
  \label{t_results}
    \begin{tabular}{lcccrrccccl}
\hline
Source Name & RA & Dec & $\sigma_\text{pos}$ & $\eta$ & $V$ & nE & nD & $S_\text{max}$ & $|{\rm V}|/{\rm I}$ & ID \\
 & (J2000) & (J2000) & (arcsec) &  &  &  &  & (mJy beam$^{-1}$) & &  \\
\hline
\multicolumn{1}{c}{Pulsars} \\
\hline
VAST J025556.2$-$530421 & 02:55:56.3 & $-$53:04:21 & 0.4 & 2997.26 & 1.40 & 7 & 7 & 52.6$\pm$0.4 & $<0.01$ & PSR J0255$-$5304 \\
VAST J041803.7$-$415414 & 04:18:03.8 & $-$41:54:14 & 0.4 & 14.76 & 0.80 & 9 & 4 & 4.2$\pm$0.2 & $<0.15$
 & PSR J0418$-$4154 \\
VAST J060046.5$-$575654 & 06:00:46.5 & $-$57:56:54 & 0.3 & 6.82 & 2.02 & 10 & 1 & 1.8$\pm$0.2 & $<0.29$
 & PSR J0600$-$5756 \\
VAST J203934.8$-$561709 & 20:39:34.8 & $-$56:17:10 & 0.4 & 6.22 & 1.12 & 6 & 2 & 1.9$\pm$0.2 & $<0.27$
 & PSR J2039$-$5617 \\
VAST J214435.6$-$523707 & 21:44:35.7 & $-$52:37:07 & 0.5 & 153.22 & 1.06 & 6 & 5 & 9.6$\pm$0.3 & $<0.08$
 & PSR J2144$-$5237 \\
VAST J215501.2$-$564158 & 21:55:01.3 & $-$56:41:59 & 0.4 & 5.99 & 1.11 & 7 & 1 & 1.7$\pm$0.2 & $<0.25$
 & PSR J2155$-$5641 \\
VAST J223651.8$-$552748 & 22:36:51.8 & $-$55:27:48 & 0.4 & 15.28 & 1.08 & 7 & 1 & 2.6$\pm$0.2 & $<0.20$
 & PSR J2236$-$5527 \\
\hline
\multicolumn{1}{c}{Stars} \\
\hline
VAST J033155.6$-$435914 & 03:31:55.7 & $-$43:59:15 & 0.4 & 25.14 & 2.21 & 6 & 1 & 4.2$\pm$0.3 & $0.88\pm 0.29$ & CD$-$44 1173 \\
VAST J040932.1$-$443538 & 04:09:32.2 & $-$44:35:38 & 0.3 & 9.77 & 1.55 & 10 & 1 & 2.6$\pm$0.2 & $0.93\pm 0.40$ & UPM J0409$-$4435 \\
VAST J052844.9$-$652650 & 05:28:44.9 & $-$65:26:50 & 0.4 & 56.17 & 0.56 & 11 & 11 & 7.8$\pm$0.4 & $0.60\pm 0.17$ &  V* AB Dor B \\
VAST J053328.0$-$425719 & 05:33:28.0 & $-$42:57:20 & 0.4 & 9.58 & 1.11 & 9 & 3 & 2.8$\pm$0.2 & $0.57\pm 0.24$ & SCR J0533$-$4257 \\
VAST J201949.8$-$581650 & 20:19:49.8 & $-$58:16:50 & 0.4 & 8.03 & 1.48 & 6 & 1 & 2.5$\pm$0.3 & $0.78\pm 0.36$ & LEHPM 2$-$783 \\
VAST J212217.5$-$454631 & 21:22:17.6 & $-$45:46:31 & 0.4 & 9.51 & 2.92 & 7 & 1 & 2.3$\pm$0.2 & $0.87\pm 0.39$ & UCAC3 89$-$412162 \\
VAST J224144.7$-$611933 & 22:41:44.8 & $-$61:19:33 & 0.5 & 29.89 & 0.78 & 6 & 4 & 4.4$\pm$0.3 & $0.60\pm 0.34$ & 2MASS J22414436$-$6119311 \\
\hline
\multicolumn{1}{c}{Other variables} \\
\hline
VAST J031648.1$-$625825 & 03:16:48.1 & $-$62:58:25 & 0.5 & 10.92 & 0.98 & 5 & 2 & 2.0$\pm$0.2 & $<0.35$ & --  \\
VAST J041908.3$-$472233 & 04:19:08.4 & $-$47:22:33 & 0.4 & 7.83 & 0.95 & 10 & 4 & 3.1$\pm$0.5 & $<0.47$ & --  \\
VAST J051151.0$-$623054 & 05:11:51.1 & $-$62:30:54 & 0.5 & 7.57 & 0.54 & 6 & 3 & 2.6$\pm$0.2 & $<0.21$ & --  \\
VAST J051452.1$-$440838 & 05:14:52.1 & $-$44:08:38 & 0.3 & 18.43 & 0.84 & 11 & 3 & 4.1$\pm$0.2 & $<0.11$ & --  \\
VAST J055639.5$-$614106 & 05:56:39.5 & $-$61:41:07 & 0.4 & 6.01 & 0.52 & 10 & 6 & 2.5$\pm$0.3 & $<0.24$ & --  \\
VAST J202339.9$-$561604 & 20:23:40.0 & $-$56:16:05 & 0.5 & 10.71 & 0.55 & 6 & 5 & 2.9$\pm$0.2 & $<0.18$ & --  \\
VAST J204100.8$-$500727 & 20:41:00.8 & $-$50:07:27 & 0.4 & 5.90 & 0.65 & 6 & 2 & 2.3$\pm$0.2 & $<0.27$ & --  \\
VAST J204421.6$-$495935 & 20:44:21.7 & $-$49:59:35 & 0.5 & 63.09 & 0.56 & 6 & 6 & 5.4$\pm$0.2 & $<0.14$ & --  \\
VAST J205441.9$-$401059 & 20:54:42.0 & $-$40:11:00 & 0.4 & 21.59 & 1.92 & 7 & 1 & 8.7$\pm$0.5 & $<0.09$ & -- \\
VAST J210110.3$-$451649 & 21:01:10.4 & $-$45:16:49 & 0.4 & 6.06 & 0.54 & 7 & 4 & 2.6$\pm$0.2 & $<0.24$ & --  \\
VAST J212623.6$-$461124 & 21:26:23.7 & $-$46:11:25 & 0.4 & 10.17 & 0.67 & 7 & 4 & 4.6$\pm$0.3 & $<0.15$ & --  \\
VAST J212758.6$-$470528 & 21:27:58.7 & $-$47:05:29 & 0.4 & 8.86 & 1.28 & 7 & 1 & 2.0$\pm$0.2 & $<0.24$ & --  \\
VAST J212857.7$-$395953 & 21:28:57.8 & $-$39:59:53 & 0.7 & 6.35 & 0.93 & 3 & 2 & 4.0$\pm$0.4 & $<0.56$ & --  \\
VAST J225351.7$-$630834 & 22:53:51.7 & $-$63:08:35 & 0.5 & 14.02 & 0.58 & 5 & 2 & 4.7$\pm$0.3 & $<0.41$ & --  \\
\hline
\end{tabular}

\end{table}
\end{landscape}

\subsection{Variability analysis and transient source density}
We found 28 highly variable or transient sources out of 155\,071 compact sources detected in at least one epoch of the VAST-P1 survey. This is a lower limit as we used relatively strict variability thresholds (Figure~\ref{f_variables}) and also exclude any source with neighbours within $30\,$arcsec (Section~\ref{s_analysis}).  In particular, about 2\% of the surveyed area was not searched for `transients' because it was within 30\,arcsec of an existing source, but as many as 8\% of sources were not searched for variability as they had neighbours.  Future analyses will relax these requirements and so should discover more transient and  variable sources.  The variability metrics we used in this analysis correspond to a variability of $\sim 50\%$. This implies that only an extremely small percentage ($0.02\%$) of sources above $\sim1.75$~mJy are variable at this level on timescales of a few months. This is consistent with (although much lower than) upper limits from previous studies at 1.4 GHz that showed that the fraction of variables on
timescales between minutes and years, and flux densities above 0.1 mJy is less than $\sim1\%$ \citep{bannister11a,bannister11b,thyagarajan11,mooley13} 
as summarised by \citet{mooley16}.

Of our 28 highly variable sources,  nine were detected in only a single epoch of our survey (see the sources with ${\rm nD} = 1$ in Table~\ref{t_results}) and hence would have been considered `transients' in previous works that calculated transient source densities and rates \citep[e.g.,][]{mooley16}. Of these nine single-epoch transients, seven are known pulsars or stars, suggesting that a majority of candidates detected in searches for extragalactic transients at this sensitivity will be foreground Galactic sources, even at Galactic latitudes of $|b|\approx 45\deg$ sampled here. These are easy to identify through cross-matching with existing catalogues, or, in some cases by their circularly polarised emission.

The other 19 sources were detected in multiple (2--7) epochs, but some of them could still have been considered `transients' in previous surveys due to the range of temporal scales probed.  For instance VAST~J041908.3$-$472233 has 4 detections (Figure~\ref{f_lightcurves}) but they all occur within a 15\,d period, with constraining non-detections before and after.  So for a typical synchrotron transient with timescale $>30\,$d at these frequencies (e.g., \citealt{metzger15}), it could represent a single event.  Even some sources that appear multiple times could be considered `transients' in the context of a 2-epoch transient rate.  VAST~J031648.1$-$625825 was visible, disappeared, and then reappeared 480\,d later.  For a 2-epoch survey metric it could then represent 2 independent transients.

For the purposes of comparison with previous work, we have calculated the two-epoch transient source density for extragalactic synchrotron sources, on timescales of $>30\,$d \citep{metzger15,carbone17}.  Future works will explore longer and shorter timescales as well as more detailed rates calculations.  We considered only the  unclassified sources from Table~\ref{t_results} and computed two numbers of `transients': total source pairs based on a constraining non-detection (upper limit lower than the lowest detection) combined with a detection, and unique sources, where in both cases all observations had been binned to 30\,d samples.  The former more accurately represents a 2-epoch snapshot rate, while the latter accounts for multiple detections of individual objects.  We found 20 total `transients` comprising 6 unique sources.  The equivalent sky area probed was 41\,034.6\,deg$^2$, based on unique pairs of observations binned to 30\,d samples \citep{carbone16}.  This then gives  a density of $(4.9\pm1.1) \times 10^{-4}$\,deg$^{-2}$ (total transients) or  $(1.5\pm0.6) \times 10^{-4}$\,deg$^{-2}$ (unique sources).  These are at a flux density of $>1.2$\,mJy (based on our median sensitivity and a 5$\sigma$ detection threshold).

\begin{figure*}
    \centering
    \includegraphics{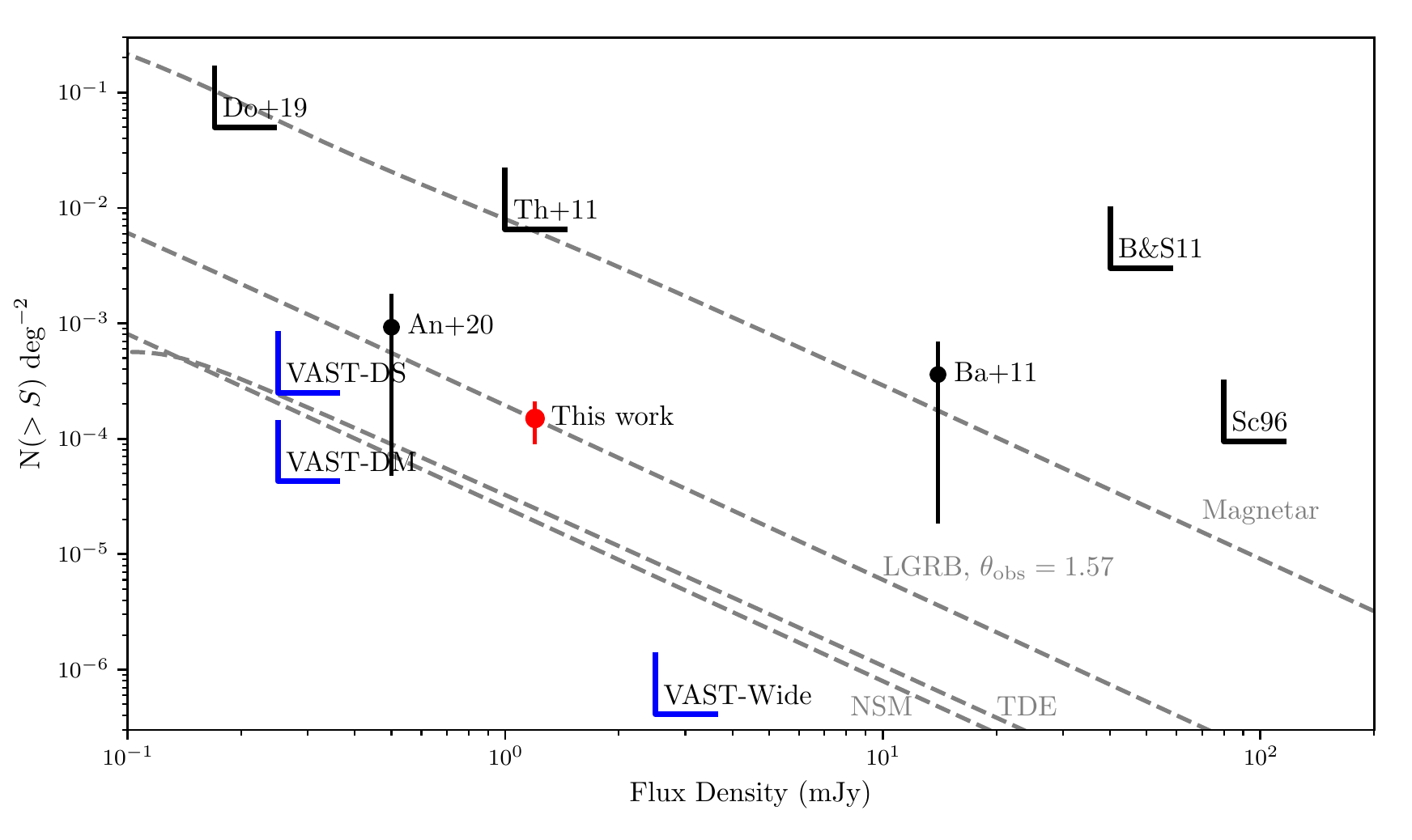}
    \caption{Two-epoch transient source surface density limits. Black wedges denote upper limits from previous searches for transients on week-month timescales \citep{2011ApJ...728L..14B,dobie19,thyagarajan11}, while black markers show rates from searches with detected transients \citep{bannister11a,anderson20} with associated two-sided 95\% confidence interval \citep{1986ApJ...303..336G}. Cyan wedges show the phase space that the proposed VAST Deep single field (VAST-DS), Deep multi-field (VAST-DM) and VAST wide surveys will be sensitive to (note that these predictions come from \citet{murphy13} and will change once the full VAST survey specifications are confirmed). The result of this work is shown in red. The predicted rates of neutron star mergers (NSM), magnetars, long and short gamma ray bursts (LGRB, SGRB) and tidal disruption events (TDE) from \citet{metzger15} are shown with dashed lines.}
    \label{f_rates}
\end{figure*}

Figure~\ref{f_rates} shows this two-epoch transient source density compared to other results (and constraining limits) from the literature\footnote{Compiled from \url{http://www.tauceti.caltech.edu/kunal/radio-transient-surveys/index.html} \citep{mooley13}.}.

\section{SUMMARY}\label{s_summary}

We have introduced the VAST Pilot Survey and described the survey strategy and observations to date. We outlined the VAST transient detection pipeline, and as a demonstration of its functionality, used it to analyse two regions (3 and 4) of the Phase I Pilot Survey.

In this analysis we found 28 highly variable and transient sources out of a total of 155\,071 sources detected in these regions. These sources included seven known pulsars, seven stars (three of which have no previous radio detections), two AGN, six galaxies and six sources yet to be identified. 

We found a transient source density of $(1.5\pm0.6) \times 10^{-4}$\,deg$^{-2}$ for a flux density limit of $>1.2$\,mJy and a 5$\sigma$ detection threshold. This is consistent with rates measured in other surveys.

Future work on the VAST pipeline will incorporate better methods for reducing the false positive rate, in particular for excluding artefacts due to nearby bright sources. We are also developing a pipeline to enable faster continuum imaging of visibility datasets (following a similar method as that of \citealt{wang21}) so that we can process data as soon as possible after the observations.

In the coming year, we will conduct the Phase II Pilot Survey, incorporating both low-band and mid-band observations. The results presented here are a significant milestone in demonstrating ASKAP's capability to detect transient sources once VAST begins full survey operations in 2022.

\begin{acknowledgements}
TM acknowledges the support of the Australian Research Council through grants FT150100099 and DP190100561.
JL and JP are supported by Australian Government Research Training Program Scholarships.
DK and AO are supported by NSF grant AST-1816492. SC acknowledges support from the NSF (AAG~1815242).
MJ was supported by NSF Physics Frontiers Center award number 1430284.
TA acknowledges support from the NSFC (12041301).
BMG acknowledges the support of the Natural Sciences and Engineering Research Council of Canada (NSERC) through grant RGPIN-2015-05948, and of the Canada Research Chairs program. GRS is supported by NSERC Discovery Grants RGPIN-2016-06569 and RGPIN-2021-04001.
SD is the recipient of an Australian Research Council Discovery Early Career Award (DE210101738).
AH acknowledges support by the I-Core Program of the Planning and Budgeting Committee and the Israel Science Foundation, and support by ISF grant 647/18. This research was supported by Grant No. 2018154 from the United States-Israel Binational Science Foundation (BSF). AH acknowledges support by the Sir Zelman Cowen Universities Fund. 

Parts of this research were conducted by the Australian Research Council Centre of Excellence for Gravitational Wave Discovery (OzGrav), project number CE170100004. This research was supported by the Sydney Informatics Hub (SIH), a core research facility at the University of Sydney.  
This work was also supported by software support resources awarded under the Astronomy Data and Computing Services (ADACS) Merit Allocation Program. ADACS is funded from the Astronomy National Collaborative Research Infrastructure Strategy (NCRIS) allocation provided by the Australian Government and managed by Astronomy Australia Limited (AAL).
The Dunlap Institute is funded through an endowment established by the David Dunlap family and the University of Toronto. 
The Australian Square Kilometre Array Pathfinder is part of the Australia Telescope National Facility which is managed by CSIRO. 
Operation of ASKAP is funded by the Australian Government with support from the National Collaborative Research Infrastructure Strategy. ASKAP uses the resources of the Pawsey Supercomputing Centre. Establishment of ASKAP, the Murchison Radio-astronomy Observatory and the Pawsey Supercomputing Centre are initiatives of the Australian Government, with support from the Government of Western Australia and the Science and Industry Endowment Fund. 

We acknowledge the Wajarri Yamatji as the traditional owners of the Murchison Radio-astronomy Observatory site. Armagh Observatory \& Planetarium is core funded by the N. Ireland Executive through the Department for Communities.
This work has made use of data from the European Space Agency (ESA) mission \gaia\ (\url{https://www.cosmos.esa.int/gaia}), processed by the {\it Gaia} Data Processing and Analysis Consortium (DPAC, \url{https://www.cosmos.esa.int/web/gaia/dpac/consortium}). Funding for the DPAC has been provided by national institutions, in particular the institutions participating in the {\it Gaia} Multilateral Agreement.
The national facility capability for SkyMapper has been funded through ARC LIEF grant LE130100104 from the Australian Research Council, awarded to the University of Sydney, the Australian National University, Swinburne University of Technology, the University of Queensland, the University of Western Australia, the University of Melbourne, Curtin University of Technology, Monash University and the Australian Astronomical Observatory. SkyMapper is owned and operated by The Australian National University's Research School of Astronomy and Astrophysics. The survey data were processed and provided by the SkyMapper Team at ANU. The SkyMapper node of the All-Sky Virtual Observatory (ASVO) is hosted at the National Computational Infrastructure (NCI). Development and support the SkyMapper node of the ASVO has been funded in part by Astronomy Australia Limited (AAL) and the Australian Government through the Commonwealth's Education Investment Fund (EIF) and National Collaborative Research Infrastructure Strategy (NCRIS), particularly the National eResearch Collaboration Tools and Resources (NeCTAR) and the Australian National Data Service Projects (ANDS).
This research made use of: Astropy,\footnote{http://www.astropy.org} a community-developed core Python package for Astronomy \citep{astropy:2013, astropy:2018}; matplotlib \citep{Hunter:2007}; and numpy \citep{harris2020array}.
\end{acknowledgements}

\input{main.bbl}

\end{document}